\documentclass{article}
\usepackage{src/arxiv}
\usepackage{graphicx} 
\usepackage{hyperref}
\usepackage{amsfonts}
\usepackage[T1]{fontenc}
\usepackage[utf8]{inputenc}

\usepackage{todonotes}
\usepackage{amsmath}
\usepackage{booktabs} 
\usepackage{setspace}
\usepackage{caption}  
\usepackage{multirow, multicol}

\title{A Rate-Distortion  Perspective on the Emergence of Number Sense in Unsupervised Generative Models}

\author{Leo D'Amato\textsuperscript{1,2}, 
Davide Nuzzi\textsuperscript{2}, 
Alberto Testolin\textsuperscript{3}, 
Ivilin Peev Stoianov\textsuperscript{2}, 
Marco Zorzi\textsuperscript{3,4*}, 
Giovanni Pezzulo\textsuperscript{2*}
\\
\bigskip
\\
\textsuperscript{\textbf{1}} Department of Control and Computer Engineering, Polytechnic University of Turin, Turin, Italy
\\
\textsuperscript{\textbf{2}} Institute of Cognitive Sciences and Technologies, National Research Council, Rome, Italy
\\
\textsuperscript{\textbf{3}} Department of General Psychology and Padova Neuroscience Center, University of Padova, Padova, Italy
\\
\textsuperscript{\textbf{4}} IRCCS San Camillo Hospital,  Venice, Italy
\\
\textsuperscript{*} Shared last authorship - correspondence: giovanni.pezzulo@istc.cnr.it, marco.zorzi@unipd.it
\\
}

\date{}

\begin{document}

\maketitle
\bigskip

\begin{abstract}

Number sense is a core cognitive ability supporting various adaptive behaviors and is foundational for mathematical learning. In this study, we investigate its emergence in unsupervised generative models through the lens of rate-distortion theory (RDT), a normative framework for understanding information processing under limited resources. We train $\beta$-Variational Autoencoders ($\beta$-VAEs) -- which embody key formal principles of RDT -- on synthetic images containing varying numbers of objects, as commonly used in numerosity perception research. We systematically vary the encoding capacity and assess the models’ sensitivity to numerosity and the robustness of the emergent numerical representations through a comprehensive set of analyses, including numerosity estimation and discrimination tasks, latent-space analysis, generative capabilities and generalization to novel stimuli. Consistent with RDT, we find that behavioral performance in numerosity perception and the ability to extract numerosity unconfounded by non-numerical visual features scale with encoding capacity according to a power law. At high capacity, the unsupervised model develops a robust neural code for numerical information, with performance closely approximating a supervised model explicitly trained for visual enumeration. It exhibits strong generative abilities and generalizes well to novel images, whereas at low capacity, the model shows marked deficits in numerosity perception and representation. Finally, comparison with human data shows that models trained at intermediate capacity levels span the full range of human behavioral performance while still developing a robust emergent numerical code. In sum, our results show that unsupervised generative models can develop a number sense and demonstrate that rate-distortion theory provides a powerful information-theoretic framework for understanding how capacity constraints shape numerosity perception.
\end{abstract}

\textbf{Keywords:} numerosity perception; rate-distortion theory, variational autoencoder; unsupervised generative model

\section{Introduction}

The ability to perceive and estimate numerosity, often referred to as "number sense," is a fundamental cognitive function observed in both humans and animals \cite{anobile2016number,dehaene2011number}. This capability enables organisms to make rapid and approximate judgments about the number of objects in their environment, facilitating adaptive behaviors such as foraging, predator avoidance, and social interactions \cite{nieder2020adaptive}. 

Empirical studies have demonstrated that numerosity perception follows well-established psychophysical principles, such as adherence to Weber’s law, which states that the just-noticeable difference in numerosity discrimination is proportional to the numerosity itself \cite{dehaene1998abstract}. Theoretical studies suggest that numerosity representations in the brain might emerge from the efficient, unsupervised coding of statistical regularities of sensory inputs, under limited computational resources \cite{cheyette2024limited,cheyette2020unified,stoianov2012emergence}. Such representations indeed emerge in neural networks that imperfectly capture the latent structure of visual scenes, highlighting that an explicit, supervised objective to count objects might not be required for learning it \cite{zorzi2018emergentist,testolin2020visual,testolin2020numerosity,creatore2021learning,verguts2004representation}. While these previous studies have provided many insights, a full mechanistic understanding of numerosity perception is still missing. For example, theoretical and empirical investigations have linked the psychophysics of numerosity perception to limited information capacity \cite{cheyette2020unified}, or more specifically to bandwidth-limited scene memory \cite{cheyette2024limited}, using abstract mathematical (Bayesian) models that cannot take real images as input. Conversely, computational studies based on unsupervised neural networks have not systematically explored the link between numerosity perception and information capacity in the information-theoretic sense \cite{dolfi2024weaker}, nor have they compared the solutions that emerge in unsupervised learning systems with those of supervised models explicitly trained to estimate the number of items in the images. 

In sum, the precise computational principles governing how numerosity representations emerge in resource-limited neural systems — and their relationship to known psychophysical laws — remain poorly understood. One open question is whether unsupervised learning alone is sufficient to develop numerical codes that generalize beyond the training set and spontaneously generate object sets in a top-down manner (i.e., numerosity production), thereby capturing the full complexity of numerosity processing, or whether such capabilities necessarily require a supervised objective explicitly tied to counting or generating numerosities. Another question concerns whether there is a systematic relation between a model’s information capacity and its ability to extract efficient numerical codes from naturalistic stimuli such as images. A further question is whether individual variability in numerosity perception (e.g., differences in "number acuity", as indexed by Weber's fraction \cite{halberda2008individual}), ranging from that of highly skilled adults to the impaired performance  in  developmental dyscalculia \cite{piazza2010developmental,dolfi2024weaker}, can be explained in terms of capacity limitations in unsupervised systems.

To address the above questions, here we investigate the emergence of number sense in unsupervised generative models through the lens of rate-distortion theory (RDT)~\cite{shannon1959coding}, a normative framework for information processing under limited resources. Originally formulated to address lossy data compression, RDT characterizes the optimal trade-off between the fidelity of information representation (distortion, $D$) and the capacity of the encoding system (rate, $R$). Formally, the goal is to minimize the distortion $D$ between an input $\boldsymbol{x}$ and its reconstruction, subject to the constraint that the amount of information encoded, $R$, does not exceed a specific capacity $C$ (i.e., $R \leq C$). This constrained optimization problem is mathematically equivalent to minimizing the Lagrangian function 
\begin{equation}
    \mathcal{L} = D + \beta (R - C),
\end{equation}
where the Lagrange multiplier $\beta$ governs the trade-off between representational accuracy and compression efficiency, leading to structured distortions in internal models of the world~\cite{d2024geometry}.

While RDT cannot be directly applied to visual image learning, a deep learning architecture known as $\beta$-Variational Autoencoder ($\beta$-VAE) approximates its principles by learning compressed latent representations of high-dimensional data under capacity constraints~\cite{d2024geometry,bates2020efficient,nagy2020optimal}. This connection is evident in the $\beta$-VAE loss function~\cite{bvae_2018,alemi2018fixing}, which functionally mirrors the RDT Lagrangian:
\begin{equation}\label{eq:bvae_loss_C}
\mathcal{L}(\theta, \phi ; \boldsymbol{x}, \boldsymbol{z}, \beta, C)=
\underbrace{ \mathbb{E}_{q_{\phi}(\boldsymbol{z} \mid \boldsymbol{x})}\left[- \log p_{\theta}(\boldsymbol{x} \mid \boldsymbol{z})\right]}_{\approx D} + 
\beta \big| \underbrace{D_{KL}\left(q_{\phi}(\boldsymbol{z} \mid \boldsymbol{x}) \| p(\boldsymbol{z})\right)}_{\approx R} - C\big|
\end{equation}

In this formulation, $\boldsymbol{x}$ denotes the high-dimensional input and $\boldsymbol{z}$ the compressed latent representation. The term $q_\phi(\boldsymbol{z}|\boldsymbol{x})$ represents the probabilistic encoder mapping inputs to the latent space, while $p_\theta(\boldsymbol{x}|\boldsymbol{z})$ is the decoder used to reconstruct the input. Both encoder and decoder are deep neural networks with  convolutional layers to effectively handle high-dimensional image data. Note that the first term in Equation \ref{eq:bvae_loss_C} (negative log-likelihood) measures the reconstruction error, corresponding to the distortion $D$, while the second term, the Kullback-Leibler divergence between the learned posterior and a fixed prior $p(\boldsymbol{z})$, acts as a proxy for the rate $R$, penalizing the complexity of the latent code. For more details about the connection between RDT and VAEs, see Section \ref{sec:met_models} and Supplementary Material \ref{suppl:sec:rdt_bvae}. Exploiting this formal link, we investigate how numerical information extracted from visual images is represented in the latent space of $\beta$-VAEs trained in an unsupervised manner to reconstruct input images. This framework allows us to assess whether the statistical structure of numerosity can be inferred without explicit supervision, how varying encoding capacity, either increasing or reducing it, leads to systematic improvements or degradations in numerosity perception, and to what extent the latent representations that emerge in an unsupervised manner align with known psychophysical laws of numerosity perception.

To assess the ability of $\beta$-VAE models trained at different capacities to learn numerosity as a summary statistic of visual scenes, we conduct a series of behavioral analyses using a dataset commonly employed in numerosity perception research, which includes images varying in numerosity, item size, and spatial arrangement. We first train a set of unsupervised $\beta$-VAEs at capacities ranging from 50 to 5000 nats, aiming to test whether a trade-off between model capacity and behavioral performance in numerosity perception tasks  emerges as predicted by RDT. We then perform detailed analyses on the numerical and generative capabilities of the unsupervised model trained at the highest capacity ($5000$ nats), aiming to test whether it develops a robust numerical code despite being not explicitly trained for this purpose. Then, we explore the effects of limited-capacity training, by analysing the numerical capabilities of the unsupervised model trained at the lowest capacity ($50$ nats). Finally, we compare the performance of unsupervised $\beta$-VAEs trained at various capacities with human behavioral data on a numerosity comparison task with images unseen during training, in order to evaluate to what extent the models reproduce the span of individual variability in human performance.

To preview our main results, we found that, in line with RDT, both behavioral performance in numerosity perception and the model's capability to extract numerical features unconfounded by other stimuli dimensions, such as size and spacing, scale with the encoding capacity of the unsupervised $\beta$-VAE. At high capacity, the model develops robust numerosity perception, closely approximating the performance of a supervised model explicitly trained for visual enumeration. The latent representation also exhibits a partial disentanglement of numerosity from other non-numerical magnitudes, enabling the model to spontaneously generate novel images with a specified number of items—a task known to be challenging even for large-scale text-to-image generative models \cite{boccato2021learning}. Moreover, the high capacity model forms a strong numerical code that generalizes to novel datasets with different statistical properties from the training data. In contrast, the lowest-capacity model shows limited generalization and impaired performance, highlighting the role of resource constraints in shaping numerosity perception. Finally, models trained at intermediate capacities can account for the range of behavioral performance and perceptual biases observed in human numerosity judgments, thereby showing that encoding capacity meaningfully captures individual variability. 

The code to reproduce the simulations, the models' checkpoints and the datasets used in this study are available at:\href{https://github.com/damat-le/num\_sense\_rdt}{https://github.com/damat-le/num\_sense\_rdt}

\section{Results}

\begin{figure}
    \centering
    \includegraphics[width=0.9\linewidth]{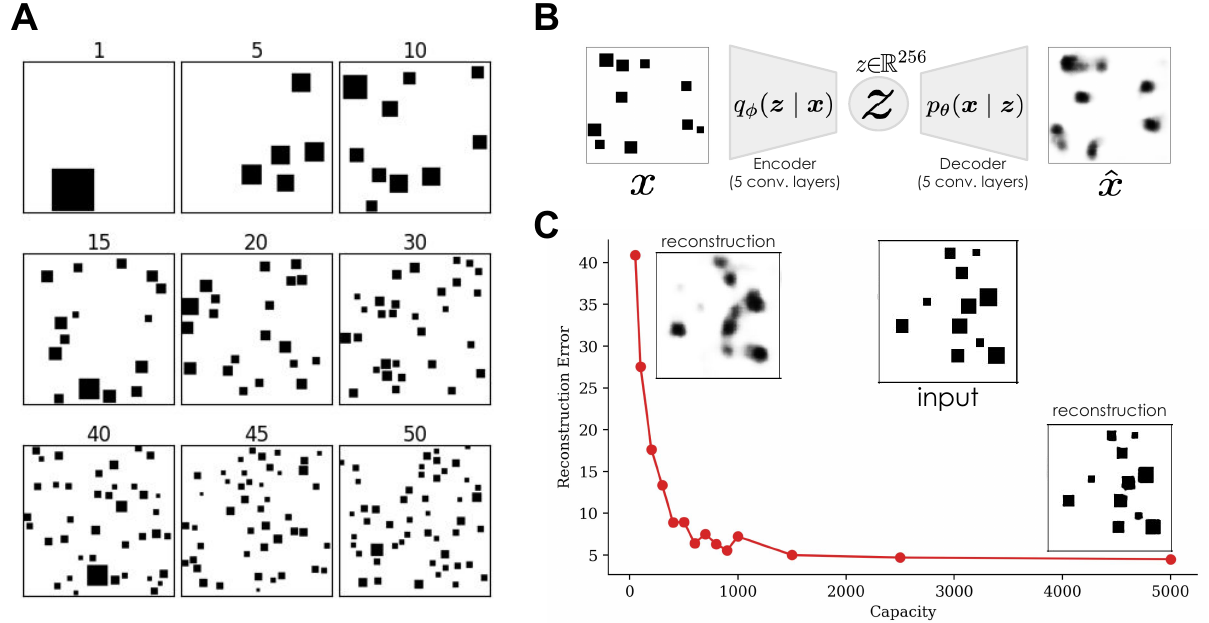}
    \caption{Dataset and computational approach. (A) Samples from the dataset used in the study. The training dataset consists of $50.000$ $256$x$256$ black-and-white images, each containing $N$ items (black squares), with $N$ ranging from $1$ to $50$. (B) Schematic illustration of the unsupervised learning model ($\beta$-VAE). (C) Reconstruction Error (mean squared error) of the $\beta$ Variational Autoencoder as a function of the encoding capacity, showing a rate-distortion trade-off. The tested capacity levels were $50$, $100$,  $200$, $300$, $400$, $500$, $600$, $700$, $800$, $900$, $1000$, $1500$, $2500$ and $5000$ nats. 
    }
    \label{fig:intro}
\end{figure}

To explore the emergence of number sense in unsupervised generative models, we trained $\beta$-Variational Autoencoders ($\beta$-VAEs) at varying encoding capacities. The training dataset comprised 50,000 black-and-white images, each containing $N$ items (black squares), with $N$ ranging from 1 to 50. Representative samples are shown in Figure \ref{fig:intro}A. 

We conducted six sets of analyses. The first set of analyses, illustrated in Section \ref{subsec:trade-off} asks if varying the encoding capacity of $\beta$-VAEs, from $50$ to $5000$ nats, produces a rate-distortion trade-off with accuracy in numerosity perception. The subsequent three analyses focus on a $\beta$-VAE trained at high encoding capacity ($5000$ nats). We refer to this model as $\beta$-VAE-5000. First, in Section \ref{subsec:behavioural_an}, we evaluate the performance of linear readouts trained on top of the model's latent representations to solve numerosity estimation and numerosity discrimination tasks. We also assess the robustness and abstractness of the learned latent representations by measuring the model's ability to generalize to a novel set of images. Second, in Section \ref{subsec:neural_an}, we explore the structure of the latent space that emerges in the model. Third, in Section \ref{subsec:generation_an}, we evaluate the model's ability to generate visual numerosities through top-down sampling from the latent state. 
In the fifth set of analyses, illustrated in Section \ref{subsec:lowcap_an}, we examine whether low-capacity training impairs numerosity perception abilities in a $\beta$-VAE model trained at low capacity (50 nats). We refer to this model as $\beta$-VAE-50. Finally, in the sixth set of analyses (Section \ref{subsec:human_comparison}), we tested $\beta$-VAE models trained at different capacities against human data, on the same set of stimuli, comparing the models’ and humans’ ability to extract and use numerosity against non-numerical visual features when comparing pairs of images.

As control simulations, we also trained supervised counterparts of the $\beta$-VAE-5000 and $\beta$-VAE-50, by augmenting the loss function with an additional term that requires to explicitly estimate the numerosity of input images (Figure \ref{fig:bvae-sup-architecture}). We refer to these supervised models as $\beta$-VAE-sup-5000 and $\beta$-VAE-sup-50, respectively. This allows to compare the latent representations of the unsupervised $\beta$-VAE-5000 and $\beta$-VAE-50 models with the numerosity-specific representations of the corresponding supervised  models, which had access to numerosity ground truth during training (Section \ref{sec:suppl_res_bvae-sup}). Technical details about model training, the dataset and the analyses are reported in Section \ref{sec:methods}.

\subsection{Rate-distortion trade-off in numerosity perception}
\label{subsec:trade-off}

The goal of this analysis is twofold. First, we aim to assess the capabilities of the unsupervised $\beta$-VAEs (Figure \ref{fig:intro}B), trained with the sole objective of reconstructing the input stimuli, in addressing phenomena related to numerosity perception. Second, we aim to explore whether a rate-distortion trade-off emerges in numerosity perception. To address these questions, we exploit the formal relation between the RTD and the encoding capacity of $\beta$-VAEs (Section \ref{sec:met_models}) and ask whether increasing or reducing the information that the model can encode about the input increases or reduces the model's numerosity perception capability. 

To this aim, we train $\beta$-Variational Autoencoders ($\beta$-VAEs) at varying encoding capacities ($50$, $100$,  $200$, $300$, $400$, $500$, $600$, $700$, $800$, $900$, $1000$, $1500$, $2500$ and $5000$ nats) on the dataset of numerical stimuli (Figure \ref{fig:intro}A).
This specific range of capacities was selected based on an information-theoretic analysis of our dataset (see Section~\ref{sec:met_dataset}), which estimates that approximately $511$ nats are required to losslessly encode the precise position and size of every item in the most complex images (N=$50$). Note that this empirical value represents the information-theoretic lower bound for an optimal discrete code and thus the capacity required for the encoding by the VAEs can be higher. Consequently, our experimental range spans from a regime of severe compression and information loss ($50$ nats) to a regime of near-perfect reconstruction ($5000$ nats), where the capacity effectively exceeds the intrinsic entropy of the data.

We then compare the models' performance in two behavioral tests widely used in human psychophysical studies \cite{piazza2010developmental,mazzocco2011impaired}. The two tests assess the model ability to represent an estimate of the number of items in a given input image (numerosity estimation) and the model ability to support the discrimination between the numerosity of items in an input image and a reference number (numerosity discrimination). These two tests provide two indices, the mean absolute error (MAE) and the Weber's fraction, which allow direct comparisons between the performance of the $\beta$-VAEs model and human number sense. Here and in the subsequent analyses the tests are performed on sets of stimuli never encountered during training, but having the same statistics. Refer to Section~\ref{sec:met_behavioural_tests} for technical details about the two behavioral tests.

\paragraph{Estimation and discrimination performance is modulated by encoding capacity.} To quantify the ability of the unsupervised model to encode numerosity information, we trained a linear regression model to predict the number of items in an image from the corresponding latent representation generated by the $\beta$-VAEs pre-trained at different capacities. The mean absolute error (MAE)  provides an overall measure of estimation accuracy for each $\beta$-VAE (see Section \ref{sec:met_behavioural_tests} for more details). To assess numerosity  discrimination, we trained a logistic regression to predict whether the input image (as encoded by its latent representation) was characterized by numerosity larger or smaller than a fixed reference number  $n_{\operatorname{ref}} = 25$. For each  $\beta$-VAE, discrimination accuracy was used to compute a psychometric function and the corresponding Weber fraction, which is a measure of precision of the internal representation (also known as number acuity) \cite{Piazza_Izard_Pinel_LeBihan_Dehaene_2004} (see Section \ref{sec:met_behavioural_tests} for more details about the computation of the Weber fraction and the training of the logistic regression).

\begin{figure}[ht!]
    \centering
    \includegraphics[width=\linewidth]{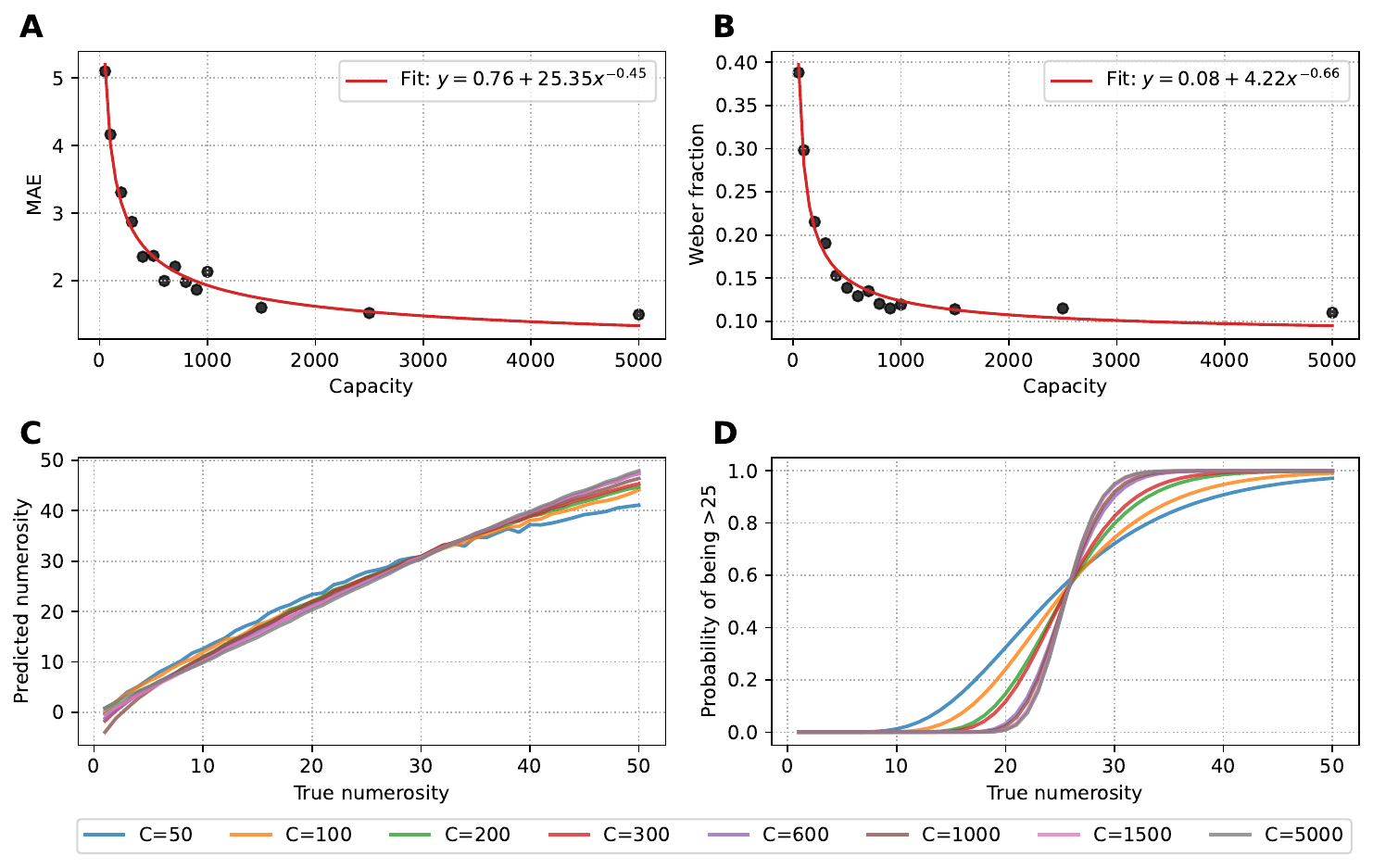}
    \caption{Rate-distortion trade-off in numerosity perception when varying the encoding capacity of $\beta$ Variational Autoencoders . (A) Mean Absolute Error (MAE) for the numerosity estimation task as a function of the encoding capacity. The best-fitting curve is obtained by a power function with a constant offset (to account for the fact that the asymptote is higher than zero). (B) Weber fraction for the numerosity discrimination task as a function of the encoding capacity. The best-fitting curve is obtained by power function with a constant offset (to account for the fact that the asymptote is higher than zero) (C) Average predictions vs True numerosity in the numerosity estimation task grouped by encoding capacity. (D) Psychometric functions obtained for the numerosity discrimination task (comparison to fixed reference $n_{\operatorname{ref}} = 25$), grouped by encoding capacity. The tested capacity levels were $50$, $100$,  $200$, $300$, $400$, $500$, $600$, $700$, $800$, $900$, $1000$, $1500$, $2500$ and $5000$ nats. 
    }
    \label{fig:num_rdcurve}
\end{figure}

The results of the analysis, summarized in Figure \ref{fig:num_rdcurve}A and \ref{fig:num_rdcurve}B, reveal a clear rate-distortion trade-off: as the encoding capacity increases, both the error in the estimation task and the Weber fraction in the discrimination task decrease. Notably, this performance-capacity trade-off follows a scaling law that is accurately captured by a power function. Therefore, our results establish a link between visual encoding capacity and numerosity perception, providing a validation of the hypothesis that performance in numerosity-related tasks is directly related to the amount of information encoded about (and hence the level of compression of) the input image (also see \cite{cheyette2024limited}). 

Further analyses of performance in the two tasks reveal capacity-dependent distortions characteristic of human numerosity perception. In the estimation task (Figure \ref{fig:num_rdcurve}C), models with limited capacity exhibit a systematic underestimation of larger numerosities, a signature consistent with behavioral studies showing that reduced encoding efficiency -- manipulated, for example, through stimulus display duration -- degrades the representation of higher numerosities \cite{cheyette2020unified}. In the discrimination task (Figure \ref{fig:num_rdcurve}D), decreasing capacity yields progressively shallower psychometric functions and enlarged Weber fractions, reflecting poorer number acuity. This pattern parallels developmental trajectories and within-group variability reported in humans, where number acuity increases sharply from childhood into adulthood but remains markedly heterogeneous across individuals \cite{halberda2008individual,piazza2010developmental}. Overall, these findings indicate that classical psychophysical signatures of numerosity perception emerge as a direct consequence of the model’s encoding capacity, consistent with the predictions of efficient coding frameworks.

\subsection{Behavioral Analyses, $\beta$-VAE-5000 model}
\label{subsec:behavioural_an}

The previous analysis showed that unsupervised $\beta$-VAEs are capable of performing numerical tasks, with their performance scaling with encoding capacity. In this analysis, we focus on the highest-capacity model ($\beta$-VAE-5000) to examine in detail how it addresses numerosity perception tasks and how it represents numerosity information.

\begin{figure}
    \centering
    \includegraphics[width=1\linewidth]{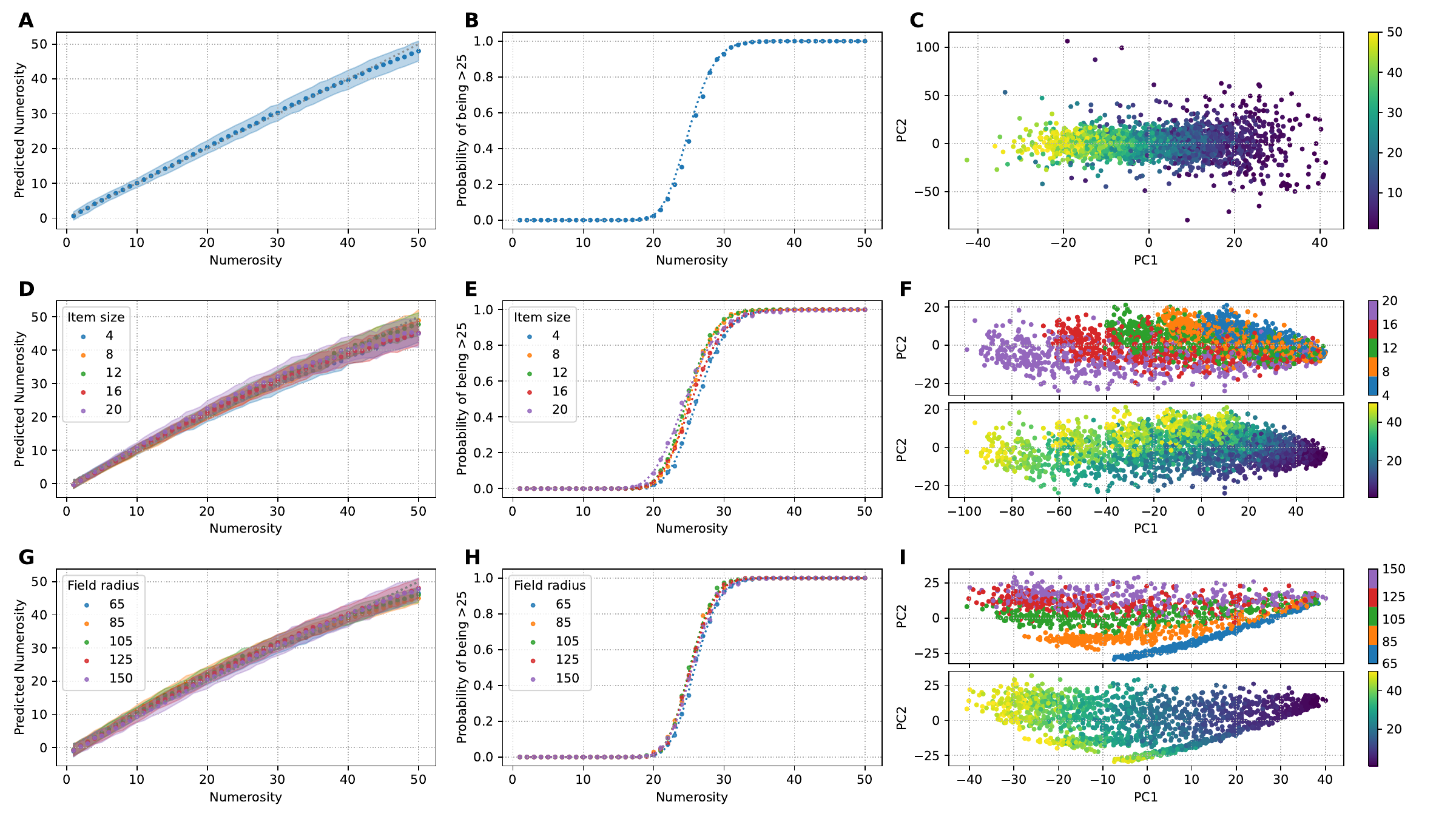}
    \caption{Analysis of the $\beta$-VAE-5000 model. (A) Results of the regression analysis simulating a numerosity estimation task. (B) Psychometric function obtained for the numerosity discrimination task. (C) Latent space visualization: first two principal components (PCs), colored by numerosity. (D-F) Generalization tests on the "item size" dataset: (D) Numerosity estimation, (E) Numerosity discrimination, (F) Latent space (PC1 vs. PC2), colored by item size (top) and numerosity (bottom). (G-I) Generalization tests on the "field radius" dataset: (G) Numerosity estimation, (H) Numerosity discrimination, (I) Latent space (PC1 vs. PC2), colored by field radius (top) and numerosity (bottom). The generalization tests are conducted by keeping the weights of the model frozen while retraining only the linear readouts. See main text for details.}
    \label{fig:bvae_retrained}
\end{figure}

\paragraph{Numerosity estimation.} Figure \ref{fig:bvae_retrained}A compares the actual numerosity of the input images (x-axis) with the numerosity predicted by the linear regression trained on the latent representations of the $\beta$-VAE-5000 (y-axis). The data points, which represent the linear regression's average predictions for each numerosity class, align closely with the diagonal, indicating strong agreement between predicted and true numerosities. As evident from Table \ref{tab:metrics_retrained}, the low mean absolute error (MAE) of regression ($\operatorname{MAE}=1.71$) confirms the model's ability to support accurate numerosity estimation. Nevertheless, the MAE exhibits a clear dependence on the number of items: it is low for small sets and increases monotonically with  numerosity  (Pearson correlation of $0.98$). This pattern of increasing error is a classic finding in  numerosity estimation \cite{cheyette2020unified,izard2008calibrating}. 

\paragraph{Numerosity discrimination.} Figure \ref{fig:bvae_retrained}B shows the probability of the logistic regression correctly discriminating whether the numerosity of an input image is larger or smaller than   the reference numerosity $n_{\operatorname{ref}}=25$. The curve is smooth and sigmoidal, with a steep transition around the reference numerosity, indicating that confidence in the discrimination is relatively high  even for numerosities close to $25$. The slope of the curve implies fine sensitivity to numerical differences in this range, supporting the model's capability for accurate numerical discrimination. As reported in Table \ref{tab:metrics_retrained}, the Weber fraction ($w=0.11$) derived from this analysis is slightly beyond what commonly reported in adult human psychophysics \cite{piazza2010developmental}.

\paragraph{Control analyses.}  To establish a performance baseline and further probe the behavioral capabilities of the $\beta$-VAE-5000 model, we conducted two control analyses. First, we evaluated the supervised $\beta$-VAE-sup-5000 model, which was explicitly trained using numerosity labels, on the same numerosity estimation and discrimination tasks. As expected, the $\beta$-VAE-sup-5000 model exhibited good performance in both tasks (Figure \ref{fig:bvae-sup_retrained}A-B, Table \ref{tab:metrics_retrained}). The performance substantially surpassed human psychophysical capabilities, as evidenced by a super-human Weber fraction as low as $w = 0.05$ in the numerosity discrimination task and a MAE of $0.80$ in the numerosity estimation task. This underscores the considerable impact of task-specific supervised training.
Second, to evaluate the reconstruction quality of the unsupervised $\beta$-VAE-5000 model and its impact on numerosity perception, we administered more challenging versions of the numerosity estimation and discrimination tasks. Instead of using the test set images as in the previous analysis, the $\beta$-VAE-5000 model first reconstructed the test images, and then performed the numerosity estimation and discrimination tasks on these reconstructed images in a zero-shot manner (i.e., without retraining any part of the model). The results demonstrate that the model performs these more challenging tasks with accuracy comparable to the original tasks (Figure \ref{fig:decoding_unsup_highc_zeroshot}), showcasing its effective reconstruction capabilities. This control analysis guarantees that the $\beta$-VAE-5000 model has correctly optimized both the rate and distortion terms during training (see Equation~\ref{eq:bvae_loss_C}).

To summarize, our behavioral analyses reveal that the unsupervised $\beta$-VAE-5000 model supports successful visual numerosity estimation and discrimination, despite being only trained to reconstruct the input. The model showed signatures of human numerosity perception, such as a greater precision in estimating smaller versus larger numerosities and its discrimination ability (measured by the Weber fraction) was slightly above the range of human psychophysics. The control analyses show that the $\beta$-VAE-5000 model approximates the performance of the supervised $\beta$-VAE-sup-5000 model -- with this latter model however showing superior performance, as expected since it was explicitly trained to count -- and has excellent reconstruction capabilities. 

\paragraph{Robustness with respect to non-numerical magnitudes.} A robust numerosity code should generalize to visual stimuli that are out-of-distribution with respect to the training data. Furthermore, it should not be confounded by additional factors, such as the size and spatial distribution of items, which often correlate with numerosity (e.g., larger and smaller items could be correlated with low and high numbers, respectively). To evaluate the generalization capability of the $\beta$-VAE-5000 model and to ensure its numerosity estimation and discrimination abilities stemmed from a robust numerosity code rather than correlated non-numerical visual features \cite{gebuis2012interplay,testolin2020visual}, we conducted the same behavioral and neural analyses on two novel datasets: "item size" and "field radius".  Crucially, all these tests were performed by keeping the weights of the $\beta$-VAE-5000 model frozen and by retraining only the linear readouts -- in order to ensure that the generalization is due to the learned internal representation of the model. 

\paragraph{Item size.} In this test, we evaluate the $\beta$-VAE-5000 model using a dataset where all items in a single image share the same size, but such size can vary across images (a condition never encountered during training). Samples from the dataset are shown in Figure \ref{fig:itemsize_dataset}. This experiment tests whether the model can estimate numerosity independently of item size and cumulative area. The size of an item in this novel dataset is given by the length (measured in pixels) of the edge of the square-shaped items. Possible edge lengths are $4, 8, 12, 16, 20$. Figure \ref{fig:bvae_retrained}D shows the predicted versus true numerosity for different item sizes. As reported in Table \ref{tab:metrics_retrained}, the model maintains strong performance, as evidenced by the small MAE values (ranging from $1.76$ to $2.65$) across all item sizes. Similarly, as evidenced by Figure \ref{fig:bvae_retrained}E, the model preserves a good number acuity since the Weber fraction remains low ($w \sim 0.12$ to $0.16$). 

\paragraph{Field radius.} In this test, we evaluate the $\beta$-VAE-5000 model using a dataset where all items share the same size and they are confined to circular fields of varying radii at the center of the image (a condition never encountered during training). Samples from the dataset are shown in Figure \ref{fig:fieldradius_dataset}. This experiment tests whether the model can estimate numerosity independently of items' spatial distribution, which corresponds to a manipulation of the density of the stimuli. Possible values of the field radius are $65, 85, 105, 125, 150$. Figure \ref{fig:bvae_retrained}G show the performance of the model in the numerosity estimation task. As reported in Table \ref{tab:metrics_retrained}, the model retains good performance, as shown by the small MAE values (ranging from $1.78$ to $2.30$) across all field radii. Similarly, as evidenced by Figure \ref{fig:bvae_retrained}H, the model preserves a good number acuity since the Weber fraction remains low ($w \sim 0.11$).

These analyses show that despite being trained only on images with randomly sized items and uniform distributions, remarkably, the model retains strong performance across a wide range of item sizes and field radii without any further update to its weights (only the linear readouts are retrained) -- indicating that the model's latent representations capture essential numerosity-related information. A similar trend is observed when testing the control $\beta$-VAE-sup-5000 model on the two new datasets (Table \ref{tab:metrics_retrained}).

\subsection{Neural encoding of numerosity, $\beta$-VAE-5000  model}
\label{subsec:neural_an}

The good performance in numerosity perception tasks suggests that the $\beta$-VAE-5000 model could develop numerosity-related features in its latent space through unsupervised learning. To test this hypothesis, we applied a Principal Component Analysis (PCA) to the 256-dimensional latent space of the model. Figure \ref{fig:bvae_retrained}C shows the first two principal components (PC1 and PC2). The points are colored according to the numerosity of the corresponding input images and range from blue (for smaller numbers) to yellow (for larger numbers). Notably, the plot reveals distinct clustering of points based on numerosity, with a gradual and smooth transition of colors along the first principal component (see Figure~\ref{fig:na_numline} for another illustration of the first PC). This result indicates that the model has successfully organized its latent representations in a way that preserves numerosity information -- as a one-dimensional, population-level "number line" -- in the first PC. Moreover, the spread of points along PC2 highlights the model's ability to encode variance in numerosity while preserving separability between different numerosities. Interestingly, the variance along PC2 is greater for smaller numbers, possibly reflecting the greater diversity of input images with lower numerosity (i.e., images containing fewer items tend to be more varied compared to those with higher numerosity). However, importantly, this increased variance does not hinder the model's ability to estimate numerosity for images with fewer items, indicating a robust numerical representation.


To obtain more insight about how numerosity information is coded in the $\beta$-VAE-5000 model's latent space, we performed two additional analyses to study whether numerosity is coded in a distributed manner across the 256 dimensions of the model's latent space or localized in a small number of dimensions. In the first analysis, we sorted the 256 neurons of the latent representation according to the Spearman correlation between their neural activity and the numerosity of the input images. We then trained 256 linear regressions, each with an increasing number of neurons, starting from those mostly correlated with numerosity, and computed the respective MAE. Figure \ref{fig:na_corr_reg}A shows that while the first 20 neurons give the greatest contribution, the MAE scores degrade gracefully, indicating that also all the other neurons contribute to a distributed numerosity coding. This result can also be appreciated by looking at the Spearman correlation score between neurons and numerosity which is higher for the first 20 neurons and degrades gracefully for the remaining ones.

In the second additional analysis, we examined the average activity of individual latent neurons in response to images containing varying numbers of items. Specifically, neural activity was first standardized using z-score normalization. Subsequently, we computed the average activity across images, grouped by numerosity. Figure \ref{fig:na_heatmap}A displays the average activity of each neuron within the latent space of the $\beta$-VAE-5000 model for each numerosity class, visualized as a heatmap. The heatmap reveals that many neurons exhibit high activity (red) when presented with images depicting small numerosities, and low activity (blue) in response to large numerosities. Furthermore, the majority of neurons span their full activity range to encode numerosities from 1 to 50, highlighting a summation coding scheme that is consistent with neuroscientific evidence \cite{roitman2007monotonic} and previous computational models \cite{dehaene1993development,stoianov2012emergence}.

Finally, we replicated all the analyses for the control $\beta$-VAE-sup-5000 model. Interestingly, the PCA analysis applied to the control $\beta$-VAE-sup-5000 model reveals a very similar structure of the latent space (Figure \ref{fig:bvae-sup_retrained}C). The supervised $\beta$-VAE-sup-5000 model also develops a more specialized neural code, in which the MAE scores degrade more abruptly (Figure \ref{fig:na_corr_reg}B). The higher Spearman correlations (in absolute value) and the higher neural activity (in absolute value) in Figure \ref{fig:na_heatmap}B also confirm this specialization. 

To summarize, this analysis the $\beta$-VAE-5000 model’s latent space reveals a structured representation of numerosity, encoded in a distributed manner across latent dimensions, with specialized neurons for small and large numbers. 

\paragraph{Disentangling of numerical and non-numerical magnitudes.} To shed further light on the robustness of the $\beta$-VAE-5000 model's numerical code, we repeated the PCA analyses, but considering the item size and field radius datasets. Interestingly, our results show a marked disentanglement of numerosity with respect to both item size (Figure \ref{fig:bvae_retrained}F) and field radius (Figure \ref{fig:bvae_retrained}I). This is evident by considering that in the item size dataset, five partially independent subspaces in PC space become apparent (color-coded in Figure \ref{fig:bvae_retrained}F, top panel), one for each of the five item sizes in the dataset. Importantly, each of these five subspaces includes data-points that span numerosities from 1 to 50 (color-coded in Figure \ref{fig:bvae_retrained}F, bottom panel). The same result is apparent when considering the field radius dataset (Figure \ref{fig:bvae_retrained}I). Furthermore, the structure of the latent space of the $\beta$-VAE-5000 model is similar to that of the $\beta$-VAE-sup-5000 control model (Figures \ref{fig:bvae-sup_retrained}F and I) but, in the latter case, a greater disentanglement of features is evident. These results show that the unsupervised model forms disentangled representations of numerosity and of other, non-numerical factors of variation, such as the size and spatial distribution of items, despite numerical and non-numerical factors correlate in the training dataset. Such disentangled latent space contributes to the robustness of the representations and also explains the good generalization results of the $\beta$-VAE-5000 model observed in the behavioral analyses in Section~\ref{subsec:behavioural_an}.

\subsection{Generative capability of the $\beta$-VAE-5000 model}
\label{subsec:generation_an}

Generative $\beta$-VAE models trained to reconstruct images also potentially acquire the capability to spontaneously generate novel images having similar statistical features. We therefore assessed the ability of the $\beta$-VAE-5000 model to generate sets with a target numerosity, by selectively activating the model's latent numerosity code. Generation is very different from simply reconstructing a specific input image (which is the model's learning objective), because it taps the model's representation of numerosity as an independently controllable visual property. This has been shown to be challenging even for state-of-the-art generative models based on transformer architectures \cite{boccato2021learning,testolin2025visual}.  

\begin{figure}[ht!]
    \centering
    \includegraphics[width=1\linewidth]{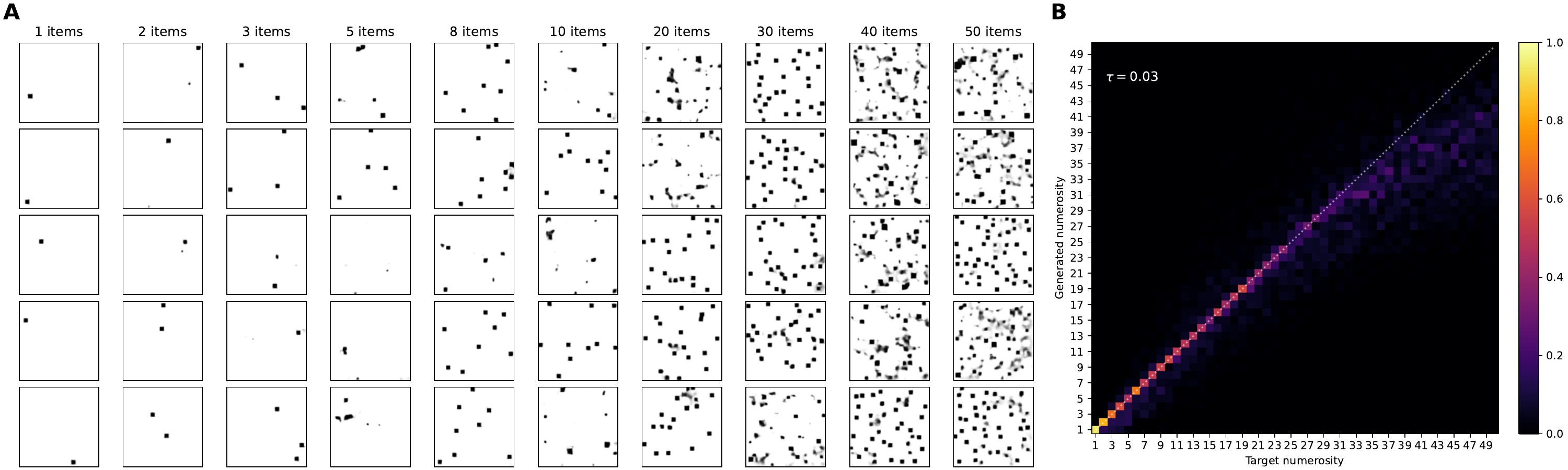}
    \caption{Generative ability of the unsupervised $\beta$-VAE-5000 model. (A) Representative generated samples for various target numerosities. (B) Confusion matrix quantifying generation accuracy in terms of the probability of generating an image with $M$ items for each target numerosity $N$ (with 100 images generated for each $N$). See Section \ref{sec:met_generative_ability} for more details. 
}
\label{fig:bvae_zeroshot_gen}
\end{figure}

To generate novel stimuli, we "prompted" the $\beta$-VAE-5000 with target numerosities, from 1 to 50, by sampling from the model's latent space (see Section \ref{sec:met_generative_ability} for a detailed description of this sampling procedure). Figure \ref{fig:bvae_zeroshot_gen}A shows that the $\beta$-VAE-5000 model tends to generate stimuli with the required numerosity, albeit (as expected) more noisy compared to the stimuli in the training set. Then, we computed the probability of generating a image with $M$ items when the target numerosity is $N$, with $M,N=1, \ldots, 50$, and we report it in Figure \ref{fig:bvae_zeroshot_gen}B. The result shows that generation accuracy is very high for stimuli with low numerosity and decreases for stimuli with high numerosity -- with the network systematically generating stimuli having a lower numerosity. These results are coherent with the other analyses of the model's behaviour in numerosity perception tasks and confirm that numerosity has been encoded as a disentangled visual feature in the model's latent representation of images.

\subsection{The effects of low capacity training in the $\beta$-VAE-50 model}
\label{subsec:lowcap_an}

\begin{figure}[ht!]
    \centering
    \includegraphics[width=1\linewidth]{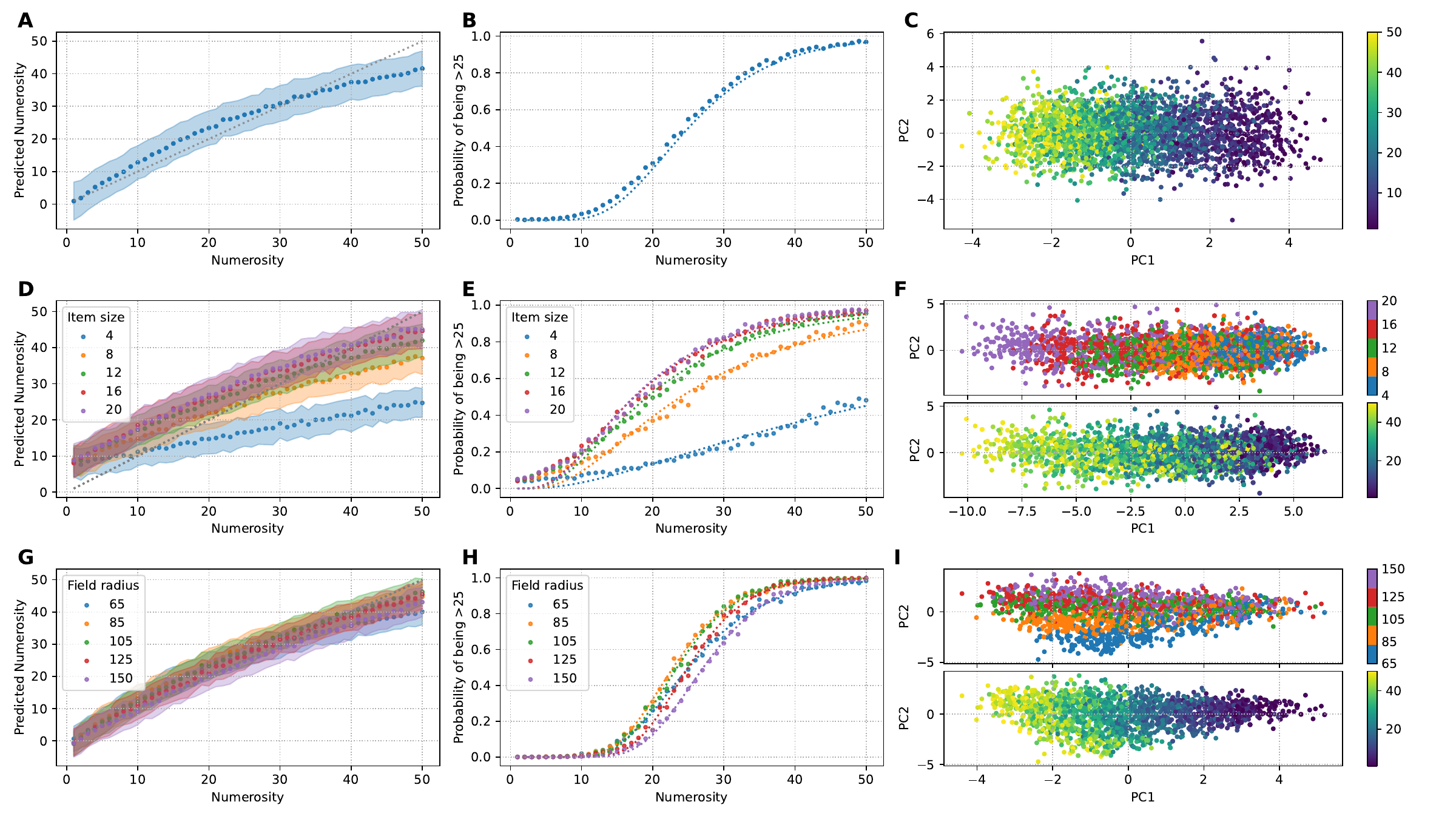}
    \caption{Analysis of the $\beta$-VAE-50 model. (A) Results of the regression analysis simulating a numerosity estimation task. (B) Psychometric function obtained for the numerosity discrimination task. (C) Latent space visualization: first two principal components (PCs), colored by numerosity. (D-F) Generalization tests on the "item size" dataset: (D) Numerosity estimation, (E) Numerosity discrimination, (F) Latent space (PC1 vs. PC2), colored by item size (top) and numerosity (bottom). (G-I) Generalization tests on the "field radius" dataset: (G) Numerosity estimation, (H) Numerosity discrimination, (I) Latent space (PC1 vs. PC2), colored by field radius (top) and numerosity (bottom). The generalization tests are conducted by keeping the weights of the model frozen while retraining only the linear readouts. See main text for details.}
\label{fig:bvae-low_retrained}
\end{figure}

So far, we focused our analysis on the $\beta$-VAE-5000 model trained at highest capacity. We now turn to the lowest capacity $\beta$-VAE ($50$ nats) to investigate how  low-capacity training influences both behavioral performance in downstream numerosity tasks and neural coding in the model  (Figure \ref{fig:bvae-low_retrained}). We refer to this model as the $\beta$-VAE-50 model. 

Compared to the high-capacity model, the low-capacity $\beta$-VAE-50 model shows poor performance in numerosity perception tasks. In the estimation task, the model's predictions show high variance and systematic underestimation of large numbers (Figure \ref{fig:bvae-low_retrained}A), indexed by $\operatorname{MAE}=5.08$. The large Weber fraction ($w = 0.39$) measured in the numerosity discrimination task also confirm the model's poor number acuity  (Figure \ref{fig:bvae-low_retrained}B). Furthermore, the latent space still appears to span a mental "number line" but is significantly more shrunk, with the range of variation of PC1 and PC2 being one order of magnitude smaller than in the $\beta$-VAE-5000 model (Figure \ref{fig:bvae-low_retrained}C). The severe capacity constraint also significantly reduces the number of neurons involved in numerosity perception ($\sim70$) compared to the $\beta$-VAE-5000 model, as evident from Figure \ref{fig:na_corr_reg}C. Finally, the $\beta$-VAE-50 shows good generalization on the "field radius" dataset (Figure \ref{fig:bvae-low_retrained}G-H) but not on the 
"item size" dataset (Figure \ref{fig:bvae-low_retrained}D-E). In fact, especially when the item size is very small, the model fails on both numerosity estimation and discrimination tasks (Table~\ref{tab:metrics_retrained}). This means that, when reducing the encoding capacity, the model is more influenced by non-numerical magnitudes that co-vary with numerosity. This is also confirmed by the  less disentangled latent spaces corresponding to the item size (Figure \ref{fig:bvae-low_retrained}F) and the field radius (Figure \ref{fig:bvae-low_retrained}I) datasets when compared to the $\beta$-VAE-5000 model.

Interestingly, a similar drastic decrease of performance (albeit less pronounced, with MAE of $3.70$ and Weber's fraction of $0.25$) and shrinking of neural representation can also be observed when a low capacity model ($\beta$-VAE-sup-50) is trained with the supervised objective to count (Figure \ref{fig:bvae-sup-low_zeroshot}). The fact that both the unsupervised and supervised model trained at low capacity learn similar neural representations is surprising, since low capacity supervised training can induce a reorganization of neural representation that prioritizes the task-relevant (here, numerical) dimensions while ignoring the others \cite{d2024geometry}. This result suggests that the decrease in performance is mainly due to capacity limitations, above and beyond the training objective, supervised or unsupervised.


In sum, our results indicate that the $\beta$-VAE-50 model trained under severe capacity constraints still shows some basic aspects of numerosity perception, such as the presence of a neural "number line". However, its ability to support numerosity tasks is significantly impaired. In particular, the limited generalization to stimuli that deviate from the training distribution in terms of non-numerical properties (e.g., item size dataset) suggests that the representation of numerosity is not sufficiently disentangled in the $\beta$-VAE-50 model and that performance in downstream tasks partly relies on non-numerical visual cues as a proxy to numerosity.

\subsection{Comparing the model's and humans' ability to extract and use numerosity features}
\label{subsec:human_comparison}

\begin{figure}
    \centering
    \includegraphics[width=\linewidth]{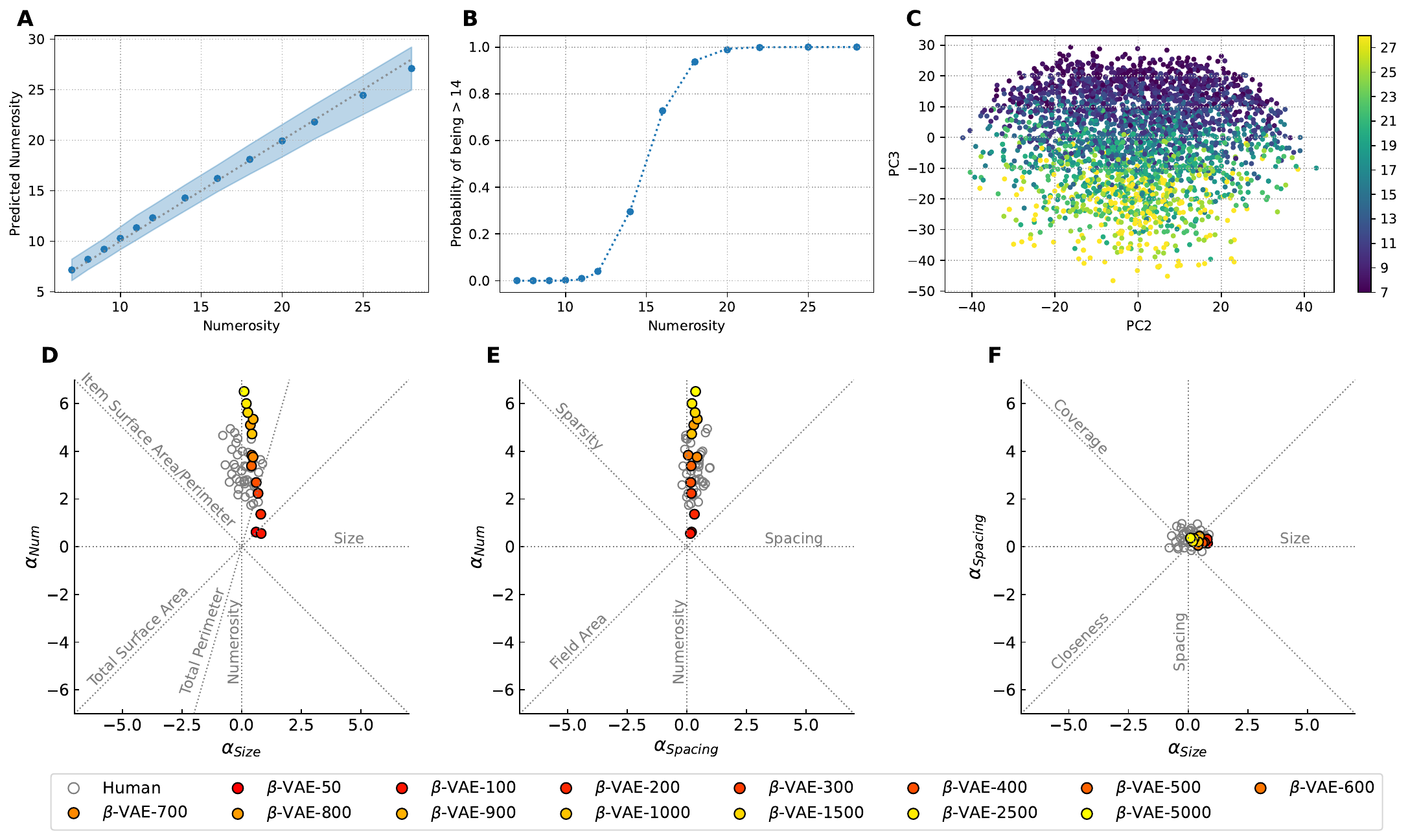}
    \caption{Comparison of $\beta$-VAE Models with Human Behavioral Data. (A-C) Performance of the unsupervised $\beta$-VAE-5000 model on the "DeWind" dataset \cite{DEWIND2015247}, evaluated by freezing the model's weights and retraining only the linear readouts. (A) Numerosity estimation task, showing predicted versus true numerosity. (B) Psychometric function for the numerosity discrimination task with reference numerosity $n_{\operatorname{ref}}=14$. (C) Visualization of the second and third principal components (PCs) of the latent space, with points colored by numerosity. (D-F) Comparison of non-numerical magnitudes influence between human participants (data from Testolin et al.~\cite{testolin2020visual}) and $\beta$-VAE models at varying encoding capacities. The plots show $\alpha$-coefficients from a Generalized Linear Model, which quantify the influence of numerosity, size, and spacing cues on discrimination judgments (technical details in Section~\ref{sec:met_humans}). Models range from low capacity ($\beta$-VAE-50) to high capacity ($\beta$-VAE-5000). (D) $\alpha_{\operatorname{Num}}$ vs. $\alpha_{\operatorname{Size}}$ coefficients. (E) $\alpha_{\operatorname{Num}}$ vs. $\alpha_{\operatorname{Spacing}}$ coefficients. (F) $\alpha_{\operatorname{Spacing}}$ vs. $\alpha_{\operatorname{Size}}$ coefficients.}
    \label{fig:human_comp}
\end{figure}

To rigorously evaluate the extent to which our models capture the nuances of human number sense, including the interplay between numerical and non-numerical (visual) magnitudes in numerosity perception (\cite{gebuis2012interplay,DEWIND2015247}), we performed a direct comparison with human behavioral data from the study of Testolin et al. \cite{testolin2020visual}. Participants were presented with a pair of images in each trial and had to judge which image had more items. Importantly, images were generated using the stimulus space of DeWind at al. \cite{DEWIND2015247}, which was specifically designed to disentangle the influence of non-numerical visual cues on numerosity perception. It therefore allows us to directly compare how numerical and non-numerical information drive the models' and human judgments.

We refer to this dataset as the "DeWind" dataset and we report some samples in Figure \ref{fig:DeWind_dataset}. We encode the same image pairs used in Testolin et al. \cite{testolin2020visual} using our frozen-weights $\beta$-VAE models trained at different capacity levels. On top of such latent representation pairs, we train a logistic regression to decide which of the two images in a pair has more dots. The decisions of the logistic regression is subsequently fitted using a Generalized Linear Model (GLM), and, for each $\beta$-VAE model, we obtain the $\alpha$-coefficients that quantify the contribution of numerosity ($\alpha_{\operatorname{Num}}$), size ($\alpha_{\operatorname{Size}}$), and spacing ($\alpha_{\operatorname{Spacing}}$) to the discrimination decisions for our $\beta$-VAE models at different encoding capacities. Refer to Section~\ref{sec:met_humans} for more details. 

The results of this analysis, shown in Figure \ref{fig:human_comp}D-F, reveal a clear pattern. Remarkably, the capability of the various $\beta$-VAE models (colored dots) to extract and use numerical features scales with their encoding capacity: the higher the encoding capacity, the more the model extracts and relies on numerical magnitude for numerosity discrimination. This result can be appreciated by noticing that the colored points are well aligned with the numerosity axis in \ref{fig:human_comp}D-E and the GLM's numerosity coefficient increases with model capacity (from red, lower capacity, to yellow, higher capacity). Furthermore, the $\beta$-VAE models (colored dots) with intermediate encoding capacity -- in the range 300-1000 nats -- are those that align well with human data from adult participants (grey dots), spanning the wide range of individual variabiliy in the strength of the numerosity coefficient. Models with higher encoding capacity are influenced by non-numerical features to a lesser extent compared to humans. Conversely, models with lower encoding capacity are influenced by non-numerical features to a greater extent compared to humans and rely more on size-related visual cues (Figure \ref{fig:human_comp}D), while the effect of spacing is not affected by capacity constraints (\ref{fig:human_comp}E). Overall, the modulating effect of encoding capacity is also resembling a human developmental pattern where age-related increase of the numerosity coefficient is associated to  decreasing weight of non-numerical cues \cite{starr2017contributions}.

In analogy with the previous analyses, we also assessed the numerosity estimation and discrimination capabilities of the $\beta$-VAE-5000 model on the "DeWind" dataset, after retraining the linear classifier but not the model's weights. As shown in Figure~\ref{fig:human_comp}A-C, the $\beta$-VAE-5000 model demonstrated excellent robustness. In the numerosity estimation task, it achieved a low Mean Absolute Error (MAE) of $1.12$, indicating high accuracy in predicting the number of items (Figure~\ref{fig:human_comp}A). In the numerosity discrimination task, the model exhibited a Weber fraction of $0.12$ (Figure~\ref{fig:human_comp}B), a value that aligns with the one observed on the training dataset and with the number acuity typically found in adult human observers \cite{piazza2010developmental}. Furthermore, the analysis of the model's latent space using PCA for this new dataset reveals that numerosity information remains systematically organized along a continuous manifold (Figure~\ref{fig:human_comp}C). These results strengthen the findings of Section \ref{subsec:behavioural_an}: in fact, the "DeWind" dataset not only has different statistics than the training dataset, but it also has items of a different geometrical shape (dots instead of squares), meaning that the numerosity representation is also abstract with respect to the shape of the items. Finally, we performed the same analysis for one of the $\beta$-VAEs ($\beta$-VAE-1000) that more closely aligns with human behavior in the "DeWind" dataset (Figure \ref{fig:bvae-1000-dewind}). We found that it shows similar performance and numerical representations as the $\beta$-VAE-5000, but with a slightly worse number acuity
(Weber fraction $w = 0.14$), which is in the range of adult human psychophysics \cite{Piazza_Izard_Pinel_LeBihan_Dehaene_2004}.

\section{Discussion}

In this study, we investigated the emergence of number sense in unsupervised generative models through the lens of rate-distortion theory (RDT, \cite{shannon1959coding}), a normative framework for understanding information processing under limited resources. We used RDT to examine whether known psychophysical phenomena related to numerosity can emerge in unsupervised generative models with bounded resources, and how capacity limitations influence both their performance in numerosity perception tasks and the structure of their numerical representations.

To this end, we trained $\beta$-Variational Autoencoders ($\beta$-VAEs) at 14 different levels of encoding capacity (ranging from 50 to 5000 nats) on a dataset commonly employed in number sense research, using an unsupervised training regime. This class of models incorporates key principles of RDT while being applicable to visual image learning. As a control, we also implemented a supervised counterpart ($\beta$-VAE-sup$)$ explicitly trained to estimate numerosity, enabling us to assess the impact of task-specific supervision on the resulting latent representations. We conducted a series of behavioral and neural analyses to evaluate the extent to which the models’ latent representations can support linear readouts in numerosity perception tasks. These analyses were designed to mirror experimental paradigms commonly used in human psychophysics—specifically, numerosity estimation and discrimination tasks—allowing for direct comparison between model performance and human perception. Linear readouts make it possible to extract numerosity information directly from the model’s latent space while minimizing potential confounds introduced by more complex, non-linear decoding architectures. Indeed, complex decoders could compensate for weak latent representations by learning intricate mappings, thereby obscuring the true nature of the encoded information. By using linear readouts, we ensure that any observed performance reflects the inherent structure and quality of the latent representations themselves.

The results of our behavioral analyses indicate that the $\beta$-VAEs can support numerosity perception tasks, despite not being explicitly trained for this purpose. Furthermore, our comparison of $\beta$-VAEs trained at 14 different encoding capacities reveals a clear rate-distortion trade-off, with model performance in numerical tasks showing power-law scaling with the amount of information that can be encoded about the input. This scaling law offers a testable prediction and it complements the results of Cheyette and Piantadosi \cite{cheyette2024limited}, who showed that individual differences in spatial memory performance (in a change localization task) were predictive of differences in a numerosity estimation task. Moreover, our numerosity comparison results show that encoding capacity is  directly linked to number acuity, as measured by the Weber fraction or the numerosity coefficient estimated using classic psychophysical models (\cite{piazza2004tuning, DEWIND2015247}). Thus, our results offer an information-theoretic account of numerosity perception based on a neural model operating directly on visual images—something not possible in previous mathematical models (e.g., \cite{cheyette2020unified,cheyette2024limited}.   

Detailed analyses of the highest capacity model ($\beta$-VAE-5000) showed a discrimination ability (Weber fraction $w = 0.11$) matching  -- or even surpassing -- the number acuity of adult human observers \cite{piazza2010developmental}, as well as strong generalization performance on novel images differing from the training data in terms of item size and spatial distribution, demonstrating the robustness and generality of the learned latent representations. Importantly, key psychophysical signatures of numerosity perception -- such as  ratio-dependent performance in numerosity comparison and scalar variability in numerosity estimation   \cite{dehaene1998abstract} --   emerged in this model (and even in the supervised counterpart explicitly trained to estimate numerosity) despite an encoding capacity exceeding the information-theoretic lower bound for optimal scene encoding. Our findings are consistent with the view that Weber scaling is the result of optimal coding principles and align with the proposal of Cheyette and Piantadosi \cite{cheyette2020unified} that enhanced sensitivity to smaller numbers reflects an optimal representation of numerosity under limited informational capacity. However, our results also reveal that the emergence of Weber's law does not require  specific capacity limits, or assumptions about the frequency distribution of numerosities (e.g., a Zipfian prior, as in the model of  \cite{cheyette2020unified}), as the latter was uniform in our training data.  

We also assessed the nature of the neural representations that emerged during the training of the unsupervised $\beta$-VAE-5000 model. Our findings reveal that, despite the absence of explicit numerosity labels during training, the model encodes robust numerical representations, largely comparable to those developed by the supervised $\beta$-VAE-sup-5000 model explicitly trained to estimate numerosity. The model shows a summation coding scheme, with the majority of neurons spanning their full activity range to encode numerosities from 1 to 50 -- consistent with empirical findings \cite{roitman2007monotonic} and previous computational models \cite{dehaene1993development,stoianov2012emergence}. The emergence of such robust numerical codes in an unsupervised system is notable, as it suggests that, first, these codes need not be innate, and second, that living organisms could learn latent codes implicitly supporting a “number sense” through statistical learning processes unrelated to numerosity perception—without (or before) having the explicit goal to count \cite{zorzi2018emergentist}. Moreover, these codes are not qualitatively different from those acquired through supervised learning with explicit labels, indicating that supervised (or “goal-driven” \cite{yamins2016using}) learning is not always necessary as a prerequisite for solving perceptual and cognitive tasks. This view aligns with growing evidence on the importance of unsupervised (pre)training in the brain \cite{zhong2025unsupervised}.

We also tested the generative capabilities of the unsupervised $\beta$-VAE-5000 model, specifically its ability to exploit the structure of its latent space to generate images with a given target numerosity. This task is particularly challenging for previous generative models based on Deep Belief Networks \cite{zorzi2018emergentist,stoianov2012emergence} and even for contemporary large-scale generative AI \cite{testolin2025visual}. Our analyses show that, in the unsupervised $\beta$-VAE-5000 model, generation accuracy is very high for stimuli with low numerosities and decreases for higher numerosities, with the network systematically generating stimuli of slightly lower numerosity than the target. These results are consistent with the model’s behavior in numerosity perception tasks, where performance is typically better for smaller numbers.

We also performed a detailed behavioral and neural analysis on a model with severe encoding capacity constraints ($\beta$-VAE-50). Our analyses show that low-capacity training significantly impairs performance in numerosity perception tasks. Furthermore, it produces neural codes for numerosity that only partially generalize to novel inputs. The latent space appears more compressed and much less disentangled, and numerosity information is encoded in a less distributed manner across neurons compared to the high-capacity $\beta$-VAE-5000, with only a few neurons correlated with numerosity and contributing to the overall representation of numerosity estimates. Interestingly, the behavior of the $\beta$-VAE-50 model recapitulates key impairments observed in developmental dyscalculia \cite{butterworth2010foundational}, which is characterized by markedly poor number acuity \cite{piazza2010developmental, dolfi2024weaker}. Though a systematic investigation is left to future studies, it is worth noting that a rate-distortion perspective suggests that impaired numerosity perception in dyscalculia can be understood as an emergent consequence of a fundamental limitation in information encoding capacity. The latter would lead to strongly compressed and noisy internal representation of numerosity, manifested as impaired precision of the Approximate Number System (ANS) \cite{wilson2007number}. 

Finally, a comparison with human behavioral data provided further evidence for the role of capacity constraints in shaping numerosity perception. By evaluating our models using the same experimental paradigm and dataset employed by Testolin et al. \cite{testolin2020visual}, we were able to quantitatively assess the influence of numerical and non-numerical features (i.e., size and spacing) on numerosity judgments. Remarkably, our analysis revealed a systematic relationship between the models’ encoding capacity and their ability to use numerosity features unconfounded by non-numerical features. The high-capacity $\beta$-VAE-5000 model developed a “purer” representation of numerosity, showing less interference from non-numerical features than human participants. Conversely, as encoding capacity was progressively reduced, the models became more sensitive to non-numerical features, more closely approximating human performance. Models with intermediate capacity ($300$-$1000$ nats) exhibited a profile that best matched the human data. These results suggest that the model's encoding capacity directly influences its capability to extract numerical features from sensory stimuli and that the characteristic perceptual biases observed in human numerosity judgments may emerge naturally from a system operating under specific information-processing constraints, consistent with the principles of rate-distortion theory.

Overall, our results  complement and extend previous findings on the neural coding of numerosity and computational models of number sense \cite{thompson2024zero, piazza2004tuning, stoianov2012emergence,cheyette2024limited,zhou2024unified}, establishing unsupervised learning as a suitable objective for developing numerosity perception and revealing a systematic relationship between encoding capacity and numerosity performance and coding. This provides a principled, information-theoretic explanation -- based on rate-distortion theory \cite{shannon1959coding} -- for both the emergence of numerical representations and their limitations under resource constraints. These findings bridge the gap between abstract theoretical accounts of efficient coding \cite{barlow1961possible}, empirical observations of neural numerosity tuning \cite{piazza2004tuning, roitman2007monotonic}, and computational models of number sense \cite{stoianov2012emergence, zorzi2018emergentist}, highlighting the sufficiency of unsupervised learning for developing human-like numerosity perception. More broadly, our findings add to a large body of literature illustrating the relevance of looking at perceptual and cognitive functions from the normative perspective of rate distortion theory and bounded rationality \cite{bates2021optimal,callaway2022rational,simon1990bounded,lieder2020resource,sims2016rate,gershman2020rationally}. The results presented here are also relevant to theories emphasizing the emergence of sophisticated cognitive abilities in generative models through unsupervised learning objectives. Previous studies have shown that inferential and predictive processing mechanisms, closely related to the variational approach of the $\beta$-VAE, can support a range of cognitive abilities, including (active) perception, planning, and cognitive control \cite{parr2022active,walsh2020evaluating,pezzulo2018hierarchical,Priorelli2023intentions}. Our study extends this line of inquiry by demonstrating that numerosity perception can also be conceptualized within this framework; however, future studies are needed to further assess the empirical validity of this claim.

In sum, by integrating insights from numerosity perception and information theory, our study aims to contribute to the understanding of how abstract numerical concepts can emerge in generative models under resource constraints. By framing these phenomena within the principles of rate-distortion theory, we demonstrate that the trade-off between encoding capacity and representational fidelity provides a principled explanation for the emergence, precision, and limitations of numerical representations. More broadly, this work sheds light on the computational principles that may underlie numerosity perception in biological systems and offers a framework for investigating the intersection between cognitive science and artificial intelligence, where information-theoretic constraints shape learning and cognition.

\section{Methods}
\label{sec:methods}

\subsection{Models}
\label{sec:met_models}
To investigate the emergence of number sense in unsupervised generative models, we employed a Beta Variational Autoencoder ($\beta$-VAE) \cite{bvae_2017}, a deep learning architecture inspired by variational inference methods in statistics. The $\beta$-VAE is particularly well-suited for studying the trade-offs between information compression and reconstruction fidelity, as its loss function has a strong formal connection to Rate-Distortion Theory (RDT) \cite{bvae_2018, park2022interpreting, shannon1959coding}. 

Historically, RDT emerged from the practical challenges of communication engineering \cite{shannon1959coding}. In this setting, a source of information, formalized as a random variable $\boldsymbol{X}$ with a probability distribution $p(\boldsymbol{x})$, sends a message to a receiver through a communication channel with limited capacity $C$. In order to fit the capacity constraints of the channel, the information content of $\boldsymbol{X}$ is encoded into a compressed representation $\boldsymbol{Z}$ via the encoding strategy described by a conditional probability distribution $q(\boldsymbol{z}|\boldsymbol{x})$. The encoder $q(\boldsymbol{z}|\boldsymbol{x})$ operates at a given rate $R(q(\boldsymbol{z}|\boldsymbol{x}))$, that quantifies the average amount of information, typically measured in bits or nats, required to represent the source $\boldsymbol{X}$ in the compressed space $\boldsymbol{Z}$. When the receiver obtains the compressed representation $\boldsymbol{Z}$, it then tries to reconstruct the original information content of $\boldsymbol{X}$ by means of a decoding strategy described by a conditional probability distribution $p(\boldsymbol{x}|\boldsymbol{z})$.
Since part of the information content of $\boldsymbol{X}$ is lost during the compression, the reconstructed content will be different with respect to the original content. This dissimilarity is measured by a distortion measure $D(q(\boldsymbol{z}|\boldsymbol{x}), p(\boldsymbol{x}|\boldsymbol{z}))$ that depends on the specific encoder-decoder pair. 
The goal is to find the optimal encoder-decoder pair that minimizes the distortion $D(q(\boldsymbol{z}|\boldsymbol{x}), p(\boldsymbol{x}|\boldsymbol{z}))$, while respecting the constraint $R(q(\boldsymbol{z}|\boldsymbol{x})) \leq C$. This corresponds to solve the following optimization problem:

 \begin{equation}
    \min_{q(\boldsymbol{z}|\boldsymbol{x}), p(\boldsymbol{x}|\boldsymbol{z})} D(q(\boldsymbol{z}|\boldsymbol{x}), p(\boldsymbol{x}|\boldsymbol{z})) \quad \text{subject to} \quad R(q(\boldsymbol{z}|\boldsymbol{x})) \leq C,
\end{equation}

that is equivalent to minimize the Lagrangian:

\begin{equation}
    \label{eq:distortion_rate_lagrangian}
        \mathcal{L} = D(q(\boldsymbol{z}|\boldsymbol{x}), p(\boldsymbol{x}|\boldsymbol{z})) + \beta \Big( R(q(\boldsymbol{z}|\boldsymbol{x})) - C \Big). 
\end{equation}

where $\beta$ is the Lagrange multiplier. However, solving this problem for high-dimensional data, such as images, is usually intractable. To overcome this limitation, we use  $\beta$-VAEs since this class of models represents an effective approximation of RDT. In fact, the loss of a $\beta$-VAE can be written as \cite{bvae_2018,alemi2018fixing}: 

\begin{equation}
\mathcal{L}(\theta, \phi ; \boldsymbol{x}, \boldsymbol{z}, \beta, C)=
\mathbb{E}_{q_{\phi}(\boldsymbol{z} \mid \boldsymbol{x})}\left[- \log p_{\theta}(\boldsymbol{x} \mid \boldsymbol{z})\right] + 
\beta \left| D_{KL}\left(q_{\phi}(\boldsymbol{z} \mid \boldsymbol{x}) \| p(\boldsymbol{z})\right) - C\right|
\end{equation}

that is functionally similar to the Lagrangian in Equation~\ref{eq:distortion_rate_lagrangian}. In fact, the term  $\mathbb{E}_{q_{\phi}(\boldsymbol{z} \mid \boldsymbol{x})}\left[- \log p_{\theta}(\boldsymbol{x} \mid \boldsymbol{z})\right]$ in Equation \ref{eq:bvae_loss_C} quantifies the discrepancy between the original input $\boldsymbol{x}$ and its reconstruction from the decoder $p_{\theta}(\boldsymbol{x} \mid \boldsymbol{z})$. This term corresponds to the distortion function $D$ in RDT and measures the fidelity of the reconstructed data. The term $D_{KL}\left(q_{\phi}(\boldsymbol{z} \mid \boldsymbol{x}) \| p(\boldsymbol{z})\right)$, instead, penalizes the deviation (in terms of Kullbach-Leibler divergence) of the learned latent distribution $q_{\phi}(\boldsymbol{z} \mid \boldsymbol{x})$ from a predefined prior distribution $p(\boldsymbol{z})$, typically a standard Gaussian. This term corresponds to the rate function $R$ in RDT, since it controls the capacity of the encoder to capture relevant information about the input.
Compared to Equation~\ref{eq:distortion_rate_lagrangian}, Equation~\ref{eq:bvae_loss_C} modifies the rate term by introducing an absolute value, thereby enforcing the constraint $R=C$ rather than the inequality $R\leq C$. This choice directly addresses the issue of posterior collapse, in which a sufficiently expressive decoder can ignore the latent variables and drive the information rate to zero \cite{alemi2018fixing}. By fixing the information rate to a nonzero target value $C$, the model is forced to actively use the latent variables, making $C$ a direct and interpretable measure of the encoding capacity (in nats).
For more details about the connection between RDT and $\beta$-VAEs and the limitations of our approach, see Supplementary Material \ref{suppl:sec:rdt_bvae}.

\paragraph{Unsupervised Model.}
The core of our analysis relied on a classical $\beta$-VAE architecture, which operates in an entirely unsupervised manner. The model learns to compress input images into a 256-dimensional latent space, capturing the essential statistical structure of the data without explicit supervision. The $\beta$-VAE consists of two main components: an encoder and a decoder. Our encoder block comprises 5 convolutional layers, each followed by a leaky ReLU activation function. The encoder maps input images to a 256-dimensional latent space, where each dimension represents a compressed feature of the input data. Our decoder block consists of 5 transposed convolutional layers, also followed by leaky ReLU activations, except for the final layer, which uses a sigmoid activation to output grayscale images. The decoder reconstructs the input images from the latent representations.

In our experiments, following the training procedure in \cite{bvae_2018}, we used the loss function in Equation \ref{eq:bvae_loss_C}. This variant allows to control the encoding capacity $C$ of the network by penalizing with a factor $\beta$ any deviation of the KL divergence from the target capacity $C$. During training, $C$ is fixed to a target value expressed in nats (ranging from 50 to 5000, see below). In addition, we used Adam optimizer with a learning rate of $10^{-3}$.

\paragraph{Supervised Model.}
We also implemented and tested a supervised version of the model, trained at the highest and lowest capacity values (50 and 5000). The supervised model incorporates an additional numerosity estimation module $f$ that is explicitly trained to predict the (ground truth) numerosity $\boldsymbol{y}$ associated to the latent vectors $\boldsymbol{z}$. The numerosity estimation module is implemented as a linear regression layer attached to the latent space of the $\beta$-VAE (Figure \ref{fig:bvae-sup-architecture}). The supervised model is thus trained using a modified loss function:
\begin{equation}
\label{eq:bvae+clf_loss}
\mathcal{L}(\theta, \phi ; \boldsymbol{x}, \boldsymbol{z}, \boldsymbol{y}, \beta)=
\mathbb{E}_{q_{\phi}(\boldsymbol{z} \mid \boldsymbol{x})}\left[- \log p_{\theta}(\boldsymbol{x} \mid \boldsymbol{z})\right] + 
\beta \left| D_{KL}\left(q_{\phi}(\boldsymbol{z} \mid \boldsymbol{x}) \| p(\boldsymbol{z})\right) - C\right| + 
\operatorname{MSE}(f(\boldsymbol{z}), \boldsymbol{y}),
\end{equation}
where $\operatorname{MSE}(f(\boldsymbol{z}), \boldsymbol{y})$ is the mean squared error of the numerosity estimation module. This joint training procedure allows the supervised model to learn numerosity-informed latent representations, while still maintaining the generative capabilities of the $\beta$-VAE. The training procedure and hyperparameters are the same as the unsupervised model.

\paragraph{Rate-distortion trade-off.} To explicitly investigate the rate-distortion trade-off, we conducted experiments varying the encoding capacity, $C$, of the $\beta$-VAE model.  Specifically, we trained multiple $\beta$-VAEs with different fixed capacity values: $50$, $100$,  $200$, $300$, $400$, $500$, $600$, $700$, $800$, $900$, $1000$, $1500$, $2500$ and $5000$ nats.  For each capacity level, we then evaluated the model's performance on both the numerosity estimation task (quantified by the Mean Absolute Error, MAE) and the numerosity discrimination task (quantified by the Weber fraction, $w$). By plotting MAE and $w$ as a function of $C$, we generated rate-distortion curves (Figures \ref{fig:num_rdcurve}A and \ref{fig:num_rdcurve}B) that visually illustrate the trade-off between compression efficiency and the accuracy of numerosity representation and perception.

\subsection{Dataset}
\label{sec:met_dataset}
All the models are trained on synthetic images that are commonly used in numerosity perception experiments. The dataset consists of 50.000 black-and-white 256x256 pixel images. Each image contains a random number of black squares, ranging from 1 to 50. Squares are placed sequentially under a hard non-overlap constraint, with their positions sampled uniformly over the image plane. Item sizes vary within each image. Square side lengths are assigned stochastically but constrained such that the cumulative surface area (i.e., the total number of black pixels) is approximately constant across different numerosities. The resulting square side length varies in the approximate range of 10-100 pixels. As a consequence, the cumulative area is uncorrelated with numerosity, preventing the models from relying on this low-level cue as a proxy for numerosity. Moreover, variability in item size precludes the use of individual item size as an alternative cue. Overall, this stimulus design isolates numerosity from confounding visual features and facilitates the investigation of robust numerical representations in the models’ latent spaces. Unless otherwise stated, all analyses reported in the main text are performed on a held-out test split with the same statistical properties as the training set.

\paragraph{Estimation of Dataset Intrinsic Entropy.}
To contextualize the encoding capacity values $C$ used in our experiments, we estimated the intrinsic information content required to losslessly describe the generated images. Since the images are constructed from discrete, procedurally generated items (black squares), the minimal information required to encode a scene is determined by the joint entropy of the items' attributes (position and size), corrected for their indistinguishability.

We estimated this value using a Monte Carlo approach. We treated a single item as a symbol defined by the tuple $(p, s)$, where $p$ represents the linearized pixel coordinate of the top-left corner and $s$ represents the side length in pixels. We generated a sample dataset of $10000$ images with $N=50$ items (the maximum numerosity condition) and computed the empirical probability distribution $\hat{P}(\text{item})$ of observing a specific position-size configuration. The Shannon entropy for a single item was computed as:
\begin{equation}
    H_{\text{item}} = - \sum \hat{P}(\text{item}) \ln \hat{P}(\text{item})
\end{equation}
For an image containing $N$ items, assuming the items are independently distributed (ignoring the minor reduction in entropy caused by the non-overlap constraint), the total entropy $H_{\text{image}}$ must account for the fact that the items are identical and interchangeable (i.e., permutation invariance). Applying the Gibbs correction for indistinguishable particles, the total estimated entropy is:
\begin{equation}
    H_{\text{image}}(N) \approx N H_{\text{item}} - \ln(N!)
\end{equation}
Our simulation yielded a single-item entropy $H_{\text{item}} \approx 13.18$ nats. For the most complex case in our dataset ($N=50$), the total required information is approximately:
\begin{equation}
    H_{\text{image}}(50) \approx 50(13.18) - \ln(50!) \approx 511 \text{ nats}
\end{equation}
This empirical value represents the information-theoretic lower bound for an optimal discrete code. However, neural networks operating with continuous latent variables and approximate posteriors generally exhibit encoding inefficiencies, requiring an operational rate higher than the theoretical minimum to achieve lossless reconstruction. Consequently, our upper bound of $C=5000$ nats provides a necessary safety margin, ensuring the model is not bottlenecked by capacity constraints and can effectively operate in a near-lossless regime. Conversely, the lower bound ($C=50$ nats) remains roughly one order of magnitude below the theoretical limit, ensuring the model operates in a regime of heavy compression.

\subsection{Behavioral Analysis}
\label{sec:met_behavioural_tests}
To evaluate the capabilities of the generative models to support numerosity perception tasks, we conducted two behavioral tests. The behavioral tests were structured to mimic experimental paradigms commonly used in human psychophysics studies \cite{piazza2010developmental,mazzocco2011impaired}, allowing for a direct comparison between the models' performance and human numerosity perception. To reproduce the two tests in our setting, we employed a linear readout protocol commonly used in representation learning \cite{bvae_2017}. This approach rests on the assumption that if the unsupervised model has learned a robust and disentangled representation of numerosity, a simple linear decoder should be able to extract this information from the latent space.

Crucially, during these behavioral tests, the weights of the pre-trained $\beta$-VAE models were frozen. The models functioned solely as feature extractors, mapping input images $\boldsymbol{x}$ to latent vectors $\boldsymbol{z}$. We then trained simple linear readouts on top of these fixed representations to solve specific psychophysical tasks. This ensures that the performance reflects the quality of the unsupervised representation rather than the capacity of a complex non-linear decoder. Moreover, all the metrics reported in the main text are computed on a test set that is different than the one used for pre-training the $\beta$-VAE models and the one used for training the linear readouts. 

\paragraph{Numerosity estimation.} 
To quantify the ability of the unsupervised model to encode  numerosity, we trained a linear regression model on top of the latent representations of the pre-trained $\beta$-VAE. This regression model predicts the number of items $N$ in the encoded images $\boldsymbol{z}$. We use the Mean Absolute Error (MAE) of regression as a measure of the overall estimation accuracy of the pre-trained $\beta$-VAE. Furthermore, for each numerosity class $N$ (ranging from 1 to 50), we computed the mean and standard deviation of the predicted estimates. This allows us to evaluate both the accuracy and uncertainty of the predictions for each numerosity class. For training the linear regression we use a different dataset than the one used for the training of the $\beta$-VAE. This dataset contains 50.000 labeled samples, 25.000 of which are used for training and 25.000 for testing. For the supervised model $\beta$-VAE-sup, we did not train a separate linear regression, as we directly used the linear module that was jointly trained with the model (Figure \ref{fig:bvae-sup-architecture}).

\paragraph{Numerosity discrimination.} To quantify the model's ability to support numerosity discrimination, we estimated the Weber fraction \cite{Norwich_1987}, a measure commonly used in numerosity perception research. In numerosity discrimination tasks, participants are typically asked to judge whether the numerosity represented in a given image is larger or smaller than a reference numerosity $n_{\operatorname{ref}}$.  In our study, we trained a logistic regression to classify latent vectors $z$ based on whether they corresponded to images with numerosity $N$ smaller (class 0) or larger (class 1) than $n_{\operatorname{ref}} = 25$. This task can be modeled using the logarithmic number line framework, where the internal representation of numerosity is assumed to be noisy and follows a Gaussian distribution on a logarithmic scale \cite{Piazza_Izard_Pinel_LeBihan_Dehaene_2004}. More specifically, the numerosity $N$ of a set of objects is represented internally by a Gaussian random variable $R \sim \mathcal{N}(\log(N), w^2)$ where the standard deviation $w$ is known as Weber fraction and quantifies the precision of the internal representation. A smaller Weber fraction indicates better acuity in numerosity discrimination. In a larger-smaller judgment task, the optimal decision strategy under a maximum likelihood criterion is to respond "larger" whenever the internal sample $R$ exceeds a criterion value $c$. The criterion $c$ represents the internal representation of the reference numerosity $n_{\operatorname{ref}}$ and it should generally coincide with the $\log(n_{\operatorname{ref}})$. 
Thus, the probability of responding "larger" to a target numerosity $N$ is given by the integral of the Gaussian distribution from $c$ to infinity:
\begin{equation}
P_{\operatorname{larger}}(N, n_{\operatorname{ref}}) = \int_{c}^{\infty} \frac{1}{\sqrt{2\pi} w} \exp\left(-\frac{(r - \log(N))^2}{2w^2}\right) dr
\end{equation}
This integral can be simplified using the error function (erf):
\begin{equation}
P_{\operatorname{larger}}(N, n_{\operatorname{ref}}) = \frac{1}{2} \left(1 + \operatorname{erf}\left(\frac{\log(N) - c}{\sqrt{2} w}\right)\right)
\end{equation}
We then fitted the model to behavioral data (logistic regression outputs) by treating $w$ and $c$ as a free parameters.
This approach allows us to directly compare the unsupervised and supervised models under varying capacity conditions.
For training the logistic regression we use a different dataset than the one used for the training of the $\beta$-VAE. This dataset contains 50.000 labeled samples (the label denotes whether the numerosity of the image is larger than 25), 25.000 of which are used for training and 25.000 for testing.

\subsection{Robustness with respect to non-numerical magnitudes}
\label{sec:met_robustness}

To assess the models' ability to generalize beyond the training distribution and to determine if they rely on spurious correlations instead of robust numerosity representation, we evaluated their behavioral and neural performance on novel datasets with varying item sizes and field radii. These datasets present conditions not encountered during training, allowing us to examine the robustness of the learned numerosity representations. Specifically, these tests by keeping the model's weights frozen while retraing only the linear readouts (linear and logistic regressions). 

\paragraph{Item Size Dataset.} We created a dataset where all items within a single image have the same size, but this size varies across different images (Figure \ref{fig:itemsize_dataset}). The size of an item (square) is given by the length of its edge measured in pixels. Possible values of item size are $4, 8, 12, 16, 20$. This contrasts with the training data, where item sizes within each image were randomly varied. This variation challenges the models to estimate numerosity independently of item size.

\paragraph{Field Radius Dataset.}  We also created a dataset where items are constrained to appear within a specific field radius measured from the center of the image (Figure \ref{fig:fieldradius_dataset}). The field radius is measured in pixels and possible values are $65, 85, 105, 125, 150$. This spatial constraint, not present in the training data (where items are uniformly distributed), probes the models' ability to estimate numerosity independently of the spatial distribution of the items.
\\

\noindent For both Item Size and Field Radius datasets, we conducted the numerosity estimation and numerosity discrimination tasks as described in Section \ref{sec:met_behavioural_tests}. More specifically, we first encoded the new images using the frozen model, then we retrained the readouts on top of these new latent representations. For each dataset, we computed the MAE and the Weber's fraction for each value of the item size / field radius. We then analyzed these performance metrics (MAE, Weber fraction) to quantify the degree of generalization achieved by each model.  Furthermore, we performed Principal Component Analysis (PCA) on the latent representations obtained from these new datasets to examine how the internal organization of the latent space adapts to the novel stimuli and whether numerosity remains disentangled from item size and field radius. The results of the unsupervised $\beta$-VAE-5000 on the new datasets are displayed in Figure \ref{fig:bvae_retrained}D-I, while the results of the $\beta$-VAE-sup-5000 model are reported in Figure \ref{fig:bvae-sup_retrained}D-I. Instead, Figure \ref{fig:bvae-low_retrained}D-I shows the results of the $\beta$-VAE-50 model.

\subsection{Reconstruction capability}
To assess the models' ability to reconstruct the input data from their latent representations, we performed a reconstruction fidelity analysis. This analysis evaluated the quality of the reconstructed images and served as a complementary assessment of the information preserved within the latent space. Specifically, we passed the images from the test set through the model's encoder to obtain their latent representations.  These latent representations were then fed into the model's decoder to generate reconstructed images. To quantify the fidelity of the reconstructions, we repeated the numerosity estimation and numerosity discrimination tasks, as detailed in Section \ref{sec:met_behavioural_tests}, using the reconstructed images as input.  This allowed us to directly compare the performance of the models on the original images with their performance on the reconstructed images.  Any significant degradation in performance on the reconstructed images would indicate a loss of relevant information during the encoding and decoding process, highlighting limitations in the reconstruction fidelity of the model.  The performance metrics obtained from these analyses (MAE, Weber fraction) were used to quantitatively evaluate the reconstruction capability of each model. The results are reported in Figure \ref{fig:decoding_unsup_highc_zeroshot}.

\subsection{Neural Analysis}
We conducted several neural analyses to examine the latent representations learned by the models. This involved:

\paragraph{Principal Component Analysis.} 
To analyze the structure of the latent space and assess the disentanglement of numerosity from other visual features, we performed Principal Component Analysis (PCA) on the 256-dimensional latent representations of the input images. To visualize the organization of the latent space, we plotted the projected latent vectors in the 2D subspace spanned by PC1 and PC2, coloring each point according to the numerosity \( N \) or other relevant features (e.g., item size or field radius). This analysis allowed us to evaluate the extent to which numerosity and other visual features are disentangled in the latent space, providing insights into the structure of the learned representations.

\paragraph{Spearman Correlation.} 
To quantify the relationship between individual neuron activations in the latent space and the numerosity of the input images, we computed the Spearman rank correlation coefficient. For each neuron \( z_i \) (\( i = 1, \ldots, 256 \)), we calculated the correlation between its activation values and the numerosity \( N \) of the corresponding input images. The Spearman correlation coefficient \( \rho_i \) for neuron \( z_i \) is defined as $\rho_i = \operatorname{Spearman}(z_i, N)$, where \( \operatorname{Spearman}(z_i, N) \) measures the strength and direction of the monotonic relationship between the neuron's activations and the numerosity. Neurons with high absolute values of \( \rho_i \) are considered particularly relevant for numerosity estimation. We report the Spearman correlation coefficients for all neurons in Figure \ref{fig:na_corr_reg}.

\paragraph{Heatmap of neural activity.} To further investigate the relationship between latent neuron activity and numerosity, we conducted an analysis to quantify the average activity of individual latent neurons in response to images containing different numbers of objects. First, we standardized the activity of each neuron across the entire dataset using z-score normalization. This ensured that all neurons were on a comparable scale, mitigating the effects of differing baseline activity levels. Subsequently, the images were grouped based on their object count (numerosity). For each numerosity class, we computed the average activity of each latent neuron across all images belonging to that class. This resulted in a matrix of average neuron activities, with rows representing individual neurons and columns representing different numerosities. This matrix was then visualized as a heatmap to reveal patterns of neuron activity as a function of numerosity, as depicted in Figure \ref{fig:na_heatmap}A. This analysis allowed us to identify neurons that exhibited preferential activation for specific numerosity ranges, providing insights into how numerosity information is encoded within the latent space of the models.

\subsection{Generative capability}
\label{sec:met_generative_ability}
To evaluate the generative performance of the models, we established a procedure for generating novel images conditioned on a specified target numerosity, without providing any input stimulus. This assessment aimed to determine whether the models could synthesize realistic and controllable images that accurately reflected the desired numerical quantity.  This involved characterizing the learned latent space representation of each numerosity and subsequently sampling from these representations to generate new visual stimuli.

\paragraph{Latent Space Characterization.}
For each target numerosity class \( N \), we modeled the distribution of latent vectors \( \boldsymbol{z} \) corresponding to images with that given numerosity using a Gaussian Mixture Model (GMM) with \( K=28 \) components.  The GMM provides a compressed probabilistic representation of the latent space for each numerosity, allowing us to sample representative latent vectors. The model is defined as:
\[
p(\boldsymbol{z} | N) = \sum_{k=1}^{K} \pi_k \mathcal{N}(\boldsymbol{z} | \boldsymbol{\mu}_k, \boldsymbol{\Sigma}_k),
\]
where \( \pi_k \) are the mixture weights, \( \boldsymbol{\mu}_k \) are the mean vectors, and \( \boldsymbol{\Sigma}_k \) are the covariance matrices for the \( k \)-th Gaussian component. The matrices \( \boldsymbol{\Sigma}_k \) are assumed to be diagonal. The parameters of the GMM (mixture weights, means, and covariances) were estimated using the Expectation-Maximization (EM) algorithm, iteratively optimizing the likelihood of the latent representations of the training images with numerosity \( N \).

\paragraph{Conditional Image Generation.}
To generate a novel image conditioned on a target numerosity \( N \), we first sample a latent vector \( \boldsymbol{z} \) from the corresponding GMM,  \( \boldsymbol{z} \sim p(\boldsymbol{z} | N) \).  The sampled latent vector \( \boldsymbol{z} \) then serves as input to the decoder network \( p_\theta(\boldsymbol{x} | \boldsymbol{z}) \), producing the generated image \( \boldsymbol{\hat{x}} \). This process allows us to synthesize images whose latent representations are statistically consistent with the specified numerosity.

\paragraph{Evaluation of Generated Images.}
To quantitatively evaluate the accuracy of the generated numerosity, we used the following procedure: given a generated image $\boldsymbol{\hat{x}}$, since it is a grayscale image, we first binarize it by applying a threshold \( \tau = 0.03 \). Pixels with intensity values less than or equal to \( \tau \) were set to black (item), while pixels with intensity values greater than \( \tau \) were set to white (background). Then we count the number of connected components in the binary image. Each connected component corresponds to an item. In this way, we can count the number of items in a generated image. For each target numerosity $N$, we generated $100$ images and computed the probability of generating an image with $M$ items when the target numerosity is \( N \). We report the probabilities in a confusion matrix (see Figure \ref{fig:bvae_zeroshot_gen}B), providing a quantitative measure of the model's ability to control numerosity during generation. Note that the threshold \( \tau \) and the number of components ($K$) were determined via a grid search to maximize the average F1-score across all numerosity classes.

\subsection{Comparison to human data}
\label{sec:met_humans}
To facilitate a direct comparison between our model's performance and human behavior, we adopted the experimental paradigm and analytical methods from Testolin et al. \cite{testolin2020visual}. This involved evaluating our pre-trained $\beta$-VAE models on the same dataset presented to human participants in \cite{testolin2020visual}. We refer to this dataset as the "DeWind" dataset, since it was designed by DeWind et al. \cite{DEWIND2015247} to systematically disentangle the influence of numerosity from correlated non-numerical magnitudes.

The dataset consists of image pairs where stimuli are generated from a multi-dimensional space defined by three orthogonal dimensions: Numerosity (the number of items, ranging from 7 to 28), Size (a composite measure of total surface area and individual item area), and Spacing (a composite measure of field area and item density). By manipulating these dimensions orthogonally, the dataset allows for a precise quantification of how each feature influences numerosity judgments. The dataset is divided into train and test splits, each containing $15.200$ image pairs. Refer to \cite{DEWIND2015247,testolin2020visual} for more details about the dataset design.

For the analysis, the weights of our $\beta$-VAE models, trained at varying capacities ($50$, $100$,  $200$, $300$, $400$, $500$, $600$, $700$, $800$, $900$, $1000$, $1500$, $2500$ and $5000$ nats), were kept frozen. The models were used to generate latent representations for image pairs from the DeWind dataset. A logistic regression was then trained on top of these latent representations pairs to perform a two-alternative forced-choice (2AFC) numerosity comparison task, i.e. to decide which of the two images has more dots. More specifically, the logistic regression receives as input a 512-dimensional vector, i.e., the concatenation of the two 256-dimensional latent representations of the two images, and outputs 1 if the first image contains more dots than the second image, 0 otherwise.
The choice behavior of this classifier was subsequently analyzed using a Generalized Linear Model (GLM), mirroring the approach used for the human data in Testolin et al. \cite{testolin2020visual}. The GLM predicted the probability of choosing a stimulus as a function of the log-ratio of Numerosity, Size, and Spacing between the two stimuli in a pair. This analysis yields a set of $\alpha$ coefficients ($\alpha_{\operatorname{Num}}$, $\alpha_{\operatorname{Size}}$, $\alpha_{\operatorname{Spacing}}$) for each $\beta$-VAE model. These coefficients quantify the weight or influence of each orthogonal dimension on the discrimination decision, providing a direct measure of the $\beta$-VAE model's reliance on numerical versus non-numerical features. The resulting coefficients for the $\beta$-VAE models were then plotted alongside the human data from Testolin et al. \cite{testolin2020visual} for direct comparison, as shown in Figure~\ref{fig:human_comp}D-F.

\section*{Acknowledgments}
This research received funding from the European Union’s Horizon 2020 Framework Programme for Research and Innovation under the Specific Grant Agreements No. 952215 (TAILOR); the European Research Council under the Grant Agreement No. 820213 (ThinkAhead); the Italian National Recovery and Resilience Plan (NRRP), M4C2, funded by the European Union - NextGenerationEU (Project IR0000011, CUP B51E22000150006, “EBRAINS-Italy”; Project PE0000013, CUP B53C22003630006, "FAIR"; Project PE0000013, CUP J93C24000320007, "FAIR"; Project PE0000006, CUP J33C22002970002 “MNESYS”), the Ministry of University and Research, PRIN PNRR P20224FESY,  PRIN 20229Z7M8N, PRIN 2022EBC78W, and the National Research Council, project iForagers. The funders had no role in study design, data collection and analysis, decision to publish, or preparation of the manuscript. We used a Generative AI model to correct typographical errors and edit language for clarity. 

\newpage


\bibliographystyle{unsrt}
\bibliography{src/references}

\newpage
\appendix
\setcounter{figure}{0} 
\renewcommand{\thefigure}{S\arabic{figure}}
\renewcommand{\theHfigure}{S\arabic{figure}}
\setcounter{table}{0} 
\renewcommand{\thetable}{S\arabic{table}}
\preto\appendix{%
   \renewcommand{\sectionautorefname}{Appendix}%
   \renewcommand{\subsectionautorefname}{Appendix Subsection}%
}
\setcounter{page}{1}

\section{Supplementary Materials}
This document contains the Supplementary Materials for the paper:\\


\noindent "A Rate-Distortion Perspective on the Emergence of Number Sense in Unsupervised Generative Models"

(Leo D'Amato, Davide Nuzzi, 
Alberto Testolin, 
Ivilin Peev Stoianov, 
Marco Zorzi, 
Giovanni Pezzulo)

\subsection{Rate-Distortion Theory and Beta Variational Autoencoders}
\label{suppl:sec:rdt_bvae}
Rate-Distortion Theory (RDT) stands as a cornerstone of information theory, providing a rigorous mathematical framework for understanding the fundamental limits of data compression and the inherent trade-off between data rate and distortion.  Its genesis can be traced back to the seminal work \cite{shannon1948mathematical} of Claude Shannon in the late 1940s and 1950s, during the era of information theory and digital communication.  Shannon's groundbreaking contributions laid the groundwork for understanding the theoretical limits of reliably transmitting information across noisy channels.  Extending this foundation, Rate-Distortion Theory, formalized in Shannon's 1959 paper "Coding theorems for a discrete source with a fidelity criterion" \cite{shannon1959coding} directly addressed the problem of lossy data compression, moving beyond the realm of perfect, lossless transmission to consider scenarios where some level of information loss is acceptable, or even necessary, to achieve efficient encoding.

Historically, RDT emerged from the practical challenges of communication engineering, where bandwidth limitations and noise inevitably constrain the amount of information that can be transmitted from a source to a receiver over a communication channel \cite{shannon1959coding}.  However, its conceptual reach extends far beyond this initial context.  The core principles of RDT, concerning the efficient allocation of resources under constraints and the management of information fidelity, are profoundly relevant to a wide range of fields, including signal processing \cite{gersho2012vector}, image and video compression \cite{ortega1998rate}, cognitive science and neuroscience \cite{atick1992could}. Indeed, the brain, operating under severe metabolic and physical constraints, can be viewed as a prime example of a resource-limited information processing system, suggesting that principles analogous to those formalized in RDT may govern the efficiency and structure of neural representations.

At its heart, Rate-Distortion Theory seeks to answer a fundamental question: given a source of information and a constraint on the available resources for encoding this information (the ‘rate’), what is the minimum achievable distortion in the reconstructed information?  To formalize this question mathematically, RDT considers a source of information, represented by a random variable $\boldsymbol{X}$ with a probability distribution $p(\boldsymbol{x})$.  The goal is to encode this information into a compressed representation, denoted by $\boldsymbol{Z}$, and then decode this representation to obtain a reconstruction $\hat{\boldsymbol{X}}$. The encoding process is formally described by a conditional probability distribution $q(\boldsymbol{z}|\boldsymbol{x})$, representing the encoder, and the decoding process by $p(\hat{\boldsymbol{x}}|\boldsymbol{z})$, representing the decoder.

The key concepts within RDT are:

\begin{itemize}
    \item \textbf{Rate (R):}  The rate quantifies the average amount of information, typically measured in bits or nats, required to represent the source $\boldsymbol{X}$ in the compressed space $\boldsymbol{Z}$.  In the context of RDT, the rate is typically defined as the mutual information between the source $\boldsymbol{X}$ and its encoded representation $\boldsymbol{Z}$:
    \begin{equation}
    \label{eq:rate_definition}
        R(q(\boldsymbol{z}|\boldsymbol{x})) = I(\boldsymbol{X}; \boldsymbol{Z})  
    \end{equation}
    In other terms, the rate $R$ essentially measures the reduction in uncertainty about $\boldsymbol{X}$ gained by observing $\boldsymbol{Z}$.  It also can be interpreted as the average number of bits or nats needed to describe $\boldsymbol{Z}$.

    \item \textbf{Distortion (D):} The distortion measures the average loss of fidelity in the reconstruction $\hat{\boldsymbol{X}}$ compared to the original source $\boldsymbol{X}$.  This loss is quantified by a distortion function $d(\boldsymbol{x}, \hat{\boldsymbol{x}})$, which measures the discrepancy between an original source value $\boldsymbol{x}$ and its reconstruction $\hat{\boldsymbol{x}}$.  The average distortion is then given by:
    \begin{equation}
        D(q(\boldsymbol{z}|\boldsymbol{x}), p(\hat{\boldsymbol{x}}|\boldsymbol{z})) = \mathbb{E}_{p(\boldsymbol{x})q(\boldsymbol{z}|\boldsymbol{x})p(\hat{\boldsymbol{x}}|\boldsymbol{z})} \left[ d(\boldsymbol{x}, \hat{\boldsymbol{x}}) \right] \label{eq:distortion_definition}
    \end{equation}
    The choice of distortion function $d(\boldsymbol{x}, \hat{\boldsymbol{x}})$ as defined in Equation~\eqref{eq:distortion_definition} is crucial and depends on the specific application and the nature of the data.  Commonly used distortion measures include the squared Euclidean distance for continuous data and the Hamming distance for discrete data. For image data, pixel-wise squared error or more perceptually relevant metrics can be employed.

    \item \textbf{Distortion-Rate Function D(R):}  The central object of study in RDT can also be framed in terms of the distortion-rate function $D(R)$. This function represents the minimum achievable distortion for a given rate budget, where the rate is constrained to be less than or equal to a certain value $C$. Formally, it is defined as:
    \begin{equation}
    \label{eq:distortion_rate_function}
        D(R) = \min_{q(\boldsymbol{z}|\boldsymbol{x}), p(\hat{\boldsymbol{x}}|\boldsymbol{z})}  D(q(\boldsymbol{z}|\boldsymbol{x}), p(\hat{\boldsymbol{x}}|\boldsymbol{z})) \quad \text{subject to} \quad R(q(\boldsymbol{z}|\boldsymbol{x})) \leq C
    \end{equation}
    The distortion-rate function $D(R)$, as shown in Equation~\eqref{eq:distortion_rate_function}, thus characterizes the fundamental limit of compression from a different perspective. It defines a lower bound on the distortion for any encoder-decoder pair that operates within a given rate constraint $C$.  In other words, for a system with a limited encoding capacity $C$, RDT seeks to find the encoding and decoding strategies that minimize the unavoidable distortion, leading to the distortion-rate function $D(R)$. 
    The constrained optimization problem in Equation~\eqref{eq:distortion_rate_function} can be equivalently formulated using the Lagrangian:
    \begin{equation}
        \mathcal{L} = 
        D(q(\boldsymbol{z}|\boldsymbol{x}), p(\hat{\boldsymbol{x}}|\boldsymbol{z})) + 
        \beta \Big( R(q(\boldsymbol{z}|\boldsymbol{x})) - C \Big) 
    \end{equation}
    where $\beta$ is the Lagrange multiplier that sets the trade-off between minimizing the distortion $D(q(\boldsymbol{z}|\boldsymbol{x}), p(\hat{\boldsymbol{x}}|\boldsymbol{z}))$ and satisfying the rate constraint $I(\boldsymbol{X}; \boldsymbol{Z}) \le C$. Minimizing this Lagrangian represents the core objective of rate-distortion theory when framed in terms of the distortion-rate function.
\end{itemize}

The rate-distortion function typically exhibits a monotonically decreasing relationship: as the allowed resources $R$ increases, the minimum achievable distortion $D(R)$ decreases (Figure \ref{fig:intro}C).  This embodies the fundamental trade-off:  to achieve higher fidelity (lower distortion), more resources (higher rate) are needed.  Conversely, to reduce resource usage (lower rate), some fidelity must be sacrificed (higher distortion).  The shape of the rate-distortion function is determined by the statistical properties of the source $\boldsymbol{X}$ and the chosen distortion measure $d(\boldsymbol{x}, \hat{\boldsymbol{x}})$.

In practice, finding the exact rate-distortion function and the optimal encoder-decoder pair is often analytically intractable, especially for complex sources like images.  However, RDT provides invaluable theoretical insights and serves as a normative benchmark against which practical compression algorithms can be evaluated.  Furthermore, the conceptual framework of RDT, emphasizing the trade-off between rate and distortion, provides a powerful lens through which to understand efficient coding in diverse domains, including the study of neural representations and cognitive systems.

While Rate-Distortion Theory provides a valuable normative framework, its direct application to complex, high-dimensional data, such as images, poses significant computational challenges.  Beta-Variational Autoencoders ($\beta$-VAEs) \cite{bvae_2017} have emerged as a powerful class of generative models that offer a practical approach to approximating the principles of RDT within the domain of deep learning.

$\beta$-VAEs are a variant of Variational Autoencoders (VAEs) \cite{kingma2013vae}, which are themselves probabilistic generative models based on variational inference.  VAEs learn to map high-dimensional input data $\boldsymbol{X}$ into a lower-dimensional latent representations $\boldsymbol{Z}$ and then to generate reconstructions $\hat{\boldsymbol{X}}$ from this latent representation. The architecture of a VAE typically consists of two main neural network components:

\begin{itemize}
    \item \textbf{Encoder $q_{\phi}(\boldsymbol{z}|\boldsymbol{x})$:}  The encoder network, parameterized by $\phi$, learns to map an input data point $\boldsymbol{x}$ to a distribution in the latent space $\boldsymbol{Z}$.  In practice, this is often implemented by parameterizing a Gaussian distribution in $\boldsymbol{Z}$ with a mean and covariance that are functions of the input $\boldsymbol{x}$, i.e., $q_{\phi}(\boldsymbol{z}|\boldsymbol{x}) = \mathcal{N}(\boldsymbol{z}; \boldsymbol{\mu}_{\phi}(\boldsymbol{x}), \boldsymbol{\Sigma}_{\phi}(\boldsymbol{x}))$.  The encoder effectively compresses the input $\boldsymbol{x}$ into a probabilistic latent representation $\boldsymbol{z}$.

    \item \textbf{Decoder $p_{\theta}(\boldsymbol{x}|\boldsymbol{z})$:} The decoder network, parameterized by $\theta$, performs the inverse mapping, taking a latent vector $\boldsymbol{z}$ and generating a reconstruction $\hat{\boldsymbol{x}}$ in the input space.  Similar to the encoder, the decoder can also be probabilistic, often modeled as a Gaussian or Bernoulli distribution depending on the data type, i.e., $p_{\theta}(\boldsymbol{x}|\boldsymbol{z}) = \mathcal{N}(\boldsymbol{x}; \boldsymbol{\mu}_{\theta}(\boldsymbol{z}), \boldsymbol{\Sigma}_{\theta}(\boldsymbol{z}))$ for continuous data or $p_{\theta}(\boldsymbol{x}|\boldsymbol{z}) = \operatorname{Bernoulli}(\boldsymbol{x}; \sigma(\nu_{\theta}(\boldsymbol{z})))$ for binary data, where $\sigma$ is the sigmoid function and $\nu_{\theta}(\boldsymbol{z})$ parameterizes the Bernoulli distribution.
\end{itemize}

The standard VAE is trained by maximizing the Evidence Lower Bound (ELBO) \cite{kingma2013vae} on the log-likelihood of the data. This is equivalent to minimize the negative ELBO and it can be expressed as the following loss function:

\begin{equation}
    \mathcal{L}_{\operatorname{VAE}}(\theta, \phi; \boldsymbol{x}, \boldsymbol{z}) = \mathbb{E}_{q_{\phi}(\boldsymbol{z}|\boldsymbol{x})} [- \log p_{\theta}(\boldsymbol{x}|\boldsymbol{z})] + D_{KL}(q_{\phi}(\boldsymbol{z}|\boldsymbol{x}) || p(\boldsymbol{z})) \label{eq:vae_loss}
\end{equation}
where $p(\boldsymbol{z})$ is a prior distribution over the latent space, typically chosen to be a standard Gaussian $\mathcal{N}(0, I)$.  The first term in the ELBO, $\mathbb{E}_{q_{\phi}(\boldsymbol{z}|\boldsymbol{x})} [- \log p_{\theta}(\boldsymbol{x}|\boldsymbol{z})]$, as seen in Equation~\eqref{eq:vae_loss}, is the reconstruction term, which encourages the decoder to generate reconstructions $\hat{\boldsymbol{x}}$ that are similar to the original inputs $\boldsymbol{x}$.  This term can be seen as minimizing the distortion in the reconstruction.  The second term, $D_{KL}(q_{\phi}(\boldsymbol{z}|\boldsymbol{x}) || p(\boldsymbol{z}))$, also part of Equation~\eqref{eq:vae_loss}, is the Kullback-Leibler (KL) divergence between the encoder's posterior distribution $q_{\phi}(\boldsymbol{z}|\boldsymbol{x})$ and the prior $p(\boldsymbol{z})$. This term acts as a regularizer, encouraging the learned latent distribution to be close to the prior distribution.  In the context of RDT, this KL divergence term can be interpreted as a constraint on the rate, limiting the complexity of the latent code.

The Beta-VAE ($\beta$-VAE) extends the standard VAE by introducing a hyperparameter $\beta$ that scales the KL divergence term:

\begin{equation}
\label{eq:beta_vae_loss}
    \mathcal{L}_{\beta \operatorname{-VAE}}(\theta, \phi; \boldsymbol{x}, \boldsymbol{z}, \beta) = \underbrace{\mathbb{E}_{q_{\phi}(\boldsymbol{z}|\boldsymbol{x})} [-\log p_{\theta}(\boldsymbol{x}|\boldsymbol{z})]}_{\approx D} + \beta \underbrace{D_{KL}(q_{\phi}(\boldsymbol{z}|\boldsymbol{x}) || p(\boldsymbol{z}))}_{\approx R} 
\end{equation}

The hyperparameter $\beta$, in Equation~\eqref{eq:beta_vae_loss}, provides a crucial control over the trade-off between reconstruction accuracy and latent space regularization.  By increasing $\beta$ to values larger than $1$, we increase the weight of the KL divergence term in the loss function.  This forces the encoder to learn latent representations that are closer to the prior distribution, effectively reducing the rate or complexity of the latent code, but potentially at the cost of increased reconstruction distortion. Conversely, setting $\beta = 1$ recovers the standard VAE loss, while $\beta < 1$ would prioritize reconstruction over latent space regularization, although this is less commonly explored in the context of disentanglement or efficient coding.

The connection between the $\beta$-VAE loss function and Rate-Distortion Theory becomes apparent when we examine the terms in Equation~\eqref{eq:beta_vae_loss} in light of the RDT framework.  The reconstruction term, $\mathbb{E}_{q_{\phi}(\boldsymbol{z}|\boldsymbol{x})} [- \log p_{\theta}(\boldsymbol{x}|\boldsymbol{z})]$, which encourages accurate reconstruction, directly corresponds to minimizing the distortion $D$ in RDT, as defined in Equation~\eqref{eq:distortion_definition}.  Specifically, if we consider the negative log-likelihood of the data under the decoder model, $-\log p_{\theta}(\boldsymbol{x}|\boldsymbol{z})$, as a measure of distortion between the original input $\boldsymbol{x}$ and its reconstruction from latent code $\boldsymbol{z}$, then minimizing the expectation of this quantity over the encoder distribution $q_{\phi}(\boldsymbol{z}|\boldsymbol{x})$ effectively minimizes the average reconstruction distortion. This correspondence becomes even more evident when we consider the common case where the decoder $p_{\theta}(\boldsymbol{x}|\boldsymbol{z})$ is modeled as an isotropic Gaussian distribution, i.e., $p_{\theta}(\boldsymbol{x}|\boldsymbol{z}) = \mathcal{N}(\boldsymbol{x}; \boldsymbol{\mu}_{\theta}(\boldsymbol{z}), \sigma^2 I)$. In this Gaussian case, the negative log-likelihood term, assuming a fixed variance $\sigma^2$, simplifies to:
$$-\log p_{\theta}(\boldsymbol{x}|\boldsymbol{z}) = \frac{1}{2\sigma^2} ||\boldsymbol{x} - \boldsymbol{\mu}_{\theta}(\boldsymbol{z})||^2 + \text{constant}.$$
Ignoring the constant term and scaling factor $\frac{1}{2\sigma^2}$, minimizing $-\log p_{\theta}(\boldsymbol{x}|\boldsymbol{z})$ becomes equivalent to minimizing the squared Euclidean distance $||\boldsymbol{x} - \boldsymbol{\mu}_{\theta}(\boldsymbol{z})||^2$ between the input $\boldsymbol{x}$ and the mean of the decoder's output $\boldsymbol{\mu}_{\theta}(\boldsymbol{z})$, which serves as the reconstruction $\hat{\boldsymbol{x}} = \boldsymbol{\mu}_{\theta}(\boldsymbol{z})$.  This squared Euclidean distance, or Mean Squared Error (MSE), is a widely used distortion measure $d$ in Rate-Distortion Theory, particularly for continuous data.

The KL divergence term, $D_{KL}(q_{\phi}(\boldsymbol{z}|\boldsymbol{x}) || p(\boldsymbol{z}))$, which regularizes the latent space, can be seen as approximating the rate $R$ in RDT, as defined in Equation~\eqref{eq:rate_definition}.  In the context of variational inference, this term arises from the need to approximate the intractable true posterior distribution $p(\boldsymbol{z}|\boldsymbol{x})$.  VAEs employ an encoder $q_{\phi}(\boldsymbol{z}|\boldsymbol{x})$ as a variational approximation to this true posterior.  By minimizing the KL divergence between $q_{\phi}(\boldsymbol{z}|\boldsymbol{x})$ and a prior distribution $p(\boldsymbol{z})$, we are essentially constraining the complexity of the approximate posterior and, consequently, the amount of information that the latent representation $\boldsymbol{z}$ can encode about the input $\boldsymbol{x}$.  If the prior $p(\boldsymbol{z})$ is chosen to be simple, such as a standard Gaussian, then pushing the approximate posterior $q_{\phi}(\boldsymbol{z}|\boldsymbol{x})$ towards $p(\boldsymbol{z})$ limits the capacity of the latent space to deviate significantly from this simple prior, thereby implicitly limiting the rate. 

Thus, minimizing the $\beta$-VAE loss function can be interpreted as approximately solving the rate-distortion optimization problem.  By varying $\beta$, we directly manipulate the emphasis on rate versus distortion, exploring the spectrum of solutions along the rate-distortion curve.  Increasing $\beta$ prioritizes rate reduction (latent space regularization) at the potential expense of increased distortion (reconstruction error), while decreasing $\beta$ allows for lower distortion but potentially at the cost of a more complex and less efficient latent representation.

It is crucial to emphasize that the $\beta$-VAE loss function and the VAE framework as a whole provide just an approximation to the true Rate-Distortion Theory optimization.  The approximation arises from several factors: First, VAEs rely on variational inference to approximate the intractable true posterior $p(\boldsymbol{z}|\boldsymbol{x})$. Second, both the encoder $q_{\phi}(\boldsymbol{z}|\boldsymbol{x})$ and decoder $p_{\theta}(\boldsymbol{x}|\boldsymbol{z})$ are parameterized by neural networks with limited capacity and specific architectural biases. These parametric choices can constrain the space of possible encoder-decoder pairs and may not perfectly achieve the optimal solutions predicted by RDT. Third, using the negative log-likelihood as a distortion measure and the KL divergence as a rate proxy are themselves approximations. In particular, the KL divergence term in Equation~\ref{eq:beta_vae_loss} serves as a tractable upper bound on the mutual information, i.e., the rate term in rate-distortion theory (Equation~\ref{eq:rate_definition}). In fact, by considering the joint distribution over data and latent codes as $q_{\phi}(\boldsymbol{x}, \boldsymbol{z}) = p(\boldsymbol{x}) q_{\phi}(\boldsymbol{z}|\boldsymbol{x})$ and also considering the mutual information $I_q(\boldsymbol{X}; \boldsymbol{Z})$ between $\boldsymbol{X}$ and $\boldsymbol{Z}$ under $q_{\phi}(\boldsymbol{x}, \boldsymbol{z})$, it is possible to write the KL term of the $\beta$-VAE loss as:

\begin{align*}
\mathbb{E}_{p(\boldsymbol{x})} & D_{\mathrm{KL}}(q_{\phi}(\boldsymbol{z}|\boldsymbol{x}) \,\|\, p(\boldsymbol{z})) =\\
&= \mathbb{E}_{p(\boldsymbol{x})} \mathbb{E}_{q_{\phi}(\boldsymbol{z}|\boldsymbol{x})} \left[ \log \frac{q_{\phi}(\boldsymbol{z}|\boldsymbol{x})}{p(\boldsymbol{z})} \right] \\
&= \mathbb{E}_{p(\boldsymbol{x})} \mathbb{E}_{q_{\phi}(\boldsymbol{z}|\boldsymbol{x})} \left[ \log \frac{q_{\phi}(\boldsymbol{z}|\boldsymbol{x})}{q(\boldsymbol{z})} + \log \frac{q(\boldsymbol{z})}{p(\boldsymbol{z})} \right] \\
&= \mathbb{E}_{p(\boldsymbol{x})} \mathbb{E}_{q_{\phi}(\boldsymbol{z}|\boldsymbol{x})} \left[ \log \frac{q_{\phi}(\boldsymbol{z}|\boldsymbol{x})}{q(\boldsymbol{z})} \right] + \mathbb{E}_{p(\boldsymbol{x})} \mathbb{E}_{q_{\phi}(\boldsymbol{z}|\boldsymbol{x})} \left[ \log \frac{q(\boldsymbol{z})}{p(\boldsymbol{z})} \right] \\
&= \mathbb{E}_{p(\boldsymbol{x})} D_{\mathrm{KL}}(q_{\phi}(\boldsymbol{z}|\boldsymbol{x}) \,\|\, q(\boldsymbol{z})) + \mathbb{E}_{q(\boldsymbol{z})} \left[ \log \frac{q(\boldsymbol{z})}{p(\boldsymbol{z})} \right] \\
&= I_q(\boldsymbol{X}; \boldsymbol{Z}) + D_{\mathrm{KL}}(q(\boldsymbol{z}) \,\|\, p(\boldsymbol{z})) \\
& \geq I_q(\boldsymbol{X}; \boldsymbol{Z})
\end{align*}

\noindent where $q(\boldsymbol{z}) = \int p(\boldsymbol{x}) q_{\phi}(\boldsymbol{z}|\boldsymbol{x}) \, d\boldsymbol{x}$ is the marginal over the latent variables. Therefore, by minimizing the KL divergence term, the $\beta$-VAE is implicitly minimizing an upper bound on the rate. 

Despite these approximations, the $\beta$-VAE framework provides a remarkably powerful and practical tool for investigating the core principles of Rate-Distortion Theory in the context of complex data and neural network models.  By training $\beta$-VAEs and systematically varying the hyperparameter $\beta$, we can empirically explore the trade-off between rate and distortion, observe how different levels of capacity constraint shape the emergent properties of latent representations, and gain valuable insights into the principles of efficient coding that may be relevant to biological cognitive systems.

\clearpage

\subsection{Analysis of the supervised $\beta$-VAE-sup models}
\label{sec:suppl_res_bvae-sup}
To provide a comparative baseline and assess the impact of explicit supervision on numerosity representation learning, we analyzed the supervised counterpart of the $\beta$-VAE-5000 model, denoted as $\beta$-VAE-sup-5000. This model was trained with additional information about the ground truth numerosity, enabling us to examine how supervision influences the emergent latent space structure and behavioral performance. The subsequent sections detail the results obtained from this analysis, mirroring the investigations performed on the unsupervised $\beta$-VAE-5000 model. We also analyzed the supervised counterpart of the $\beta$-VAE-50 model, denoted as $\beta$-VAE-sup-50.\\


\begin{figure}[ht!]
    \centering
    \includegraphics[width=1\linewidth]{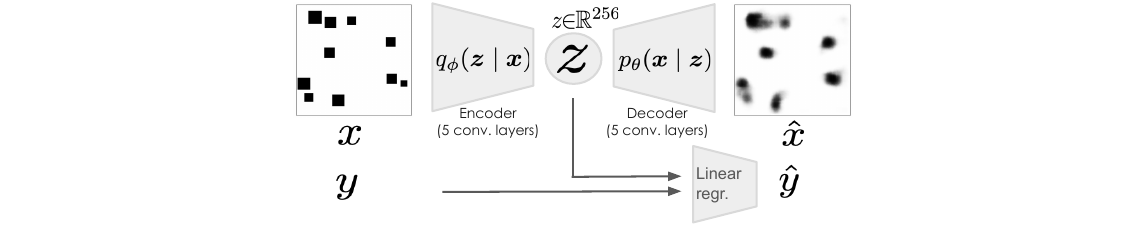}
    \caption{Architecture of the supervised $\beta$-VAE-sup model. A linear layer takes as input the latent representation $\boldsymbol{z}$ of the VAE and the corresponding ground truth numerosity $y$ to predict the number of items in the image. The linear layer is trained jointly with the VAE component by adding a MSE term in the VAE's loss function (Equation \ref{eq:bvae+clf_loss}).}
    \label{fig:bvae-sup-architecture}
\end{figure}

\begin{figure}[ht!]
    \centering
    \includegraphics[width=1\linewidth]{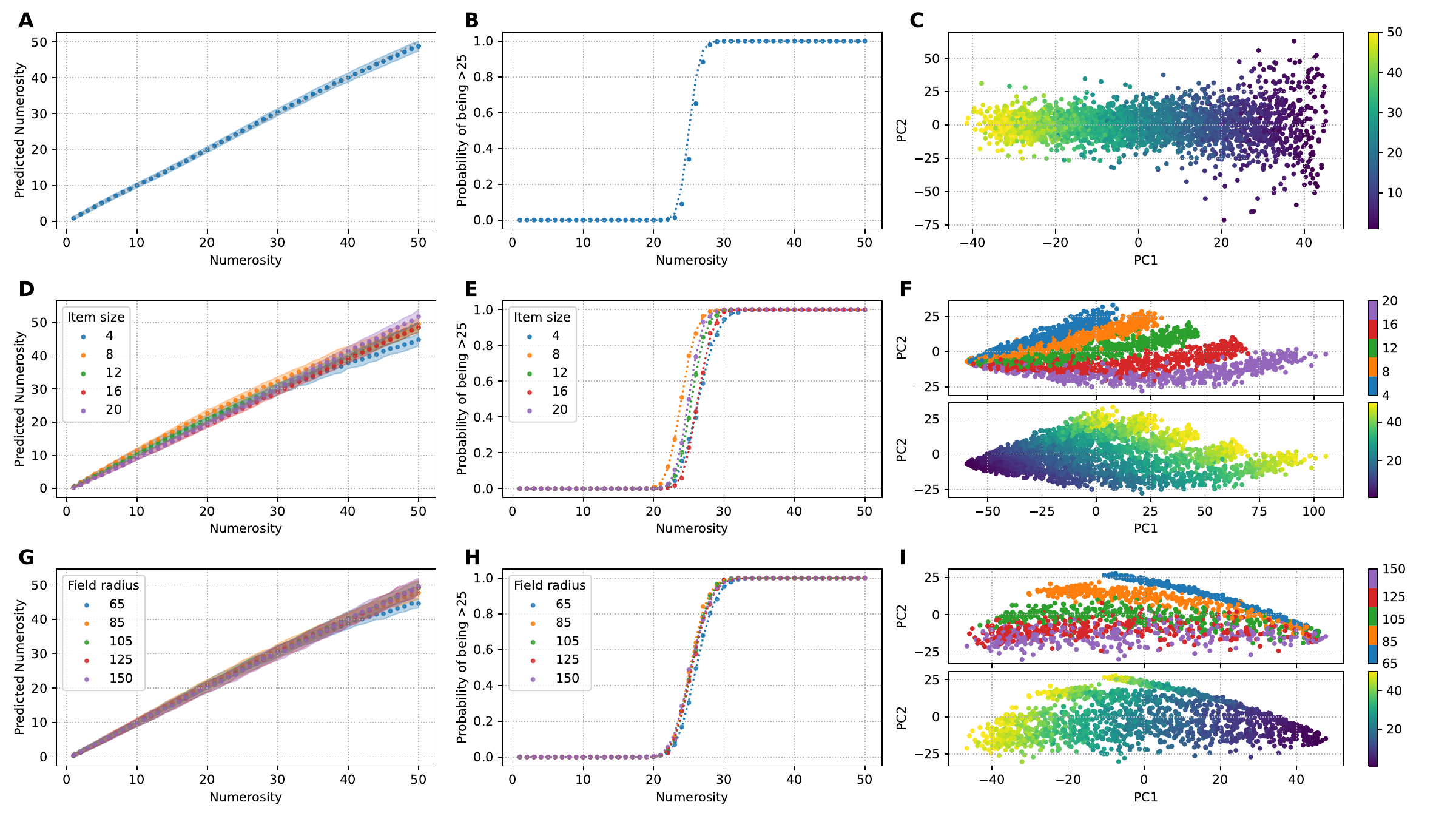}
    \caption{Analysis of the $\beta$-VAE-sup-5000 model with retrained regressions. All panels are analogous to those described in Figure \ref{fig:bvae_retrained}. The generalization tests in panels D, E, G and H are conducted without by keeping the model's weights frozen while retraining only the readouts. This allows comparison of the model's performance in numerosity estimation and discrimination tasks and the organization of its latent space under supervised training conditions.}
    \label{fig:bvae-sup_retrained}
\end{figure}

\begin{figure}[ht!]
    \centering
    \includegraphics[width=1\linewidth]{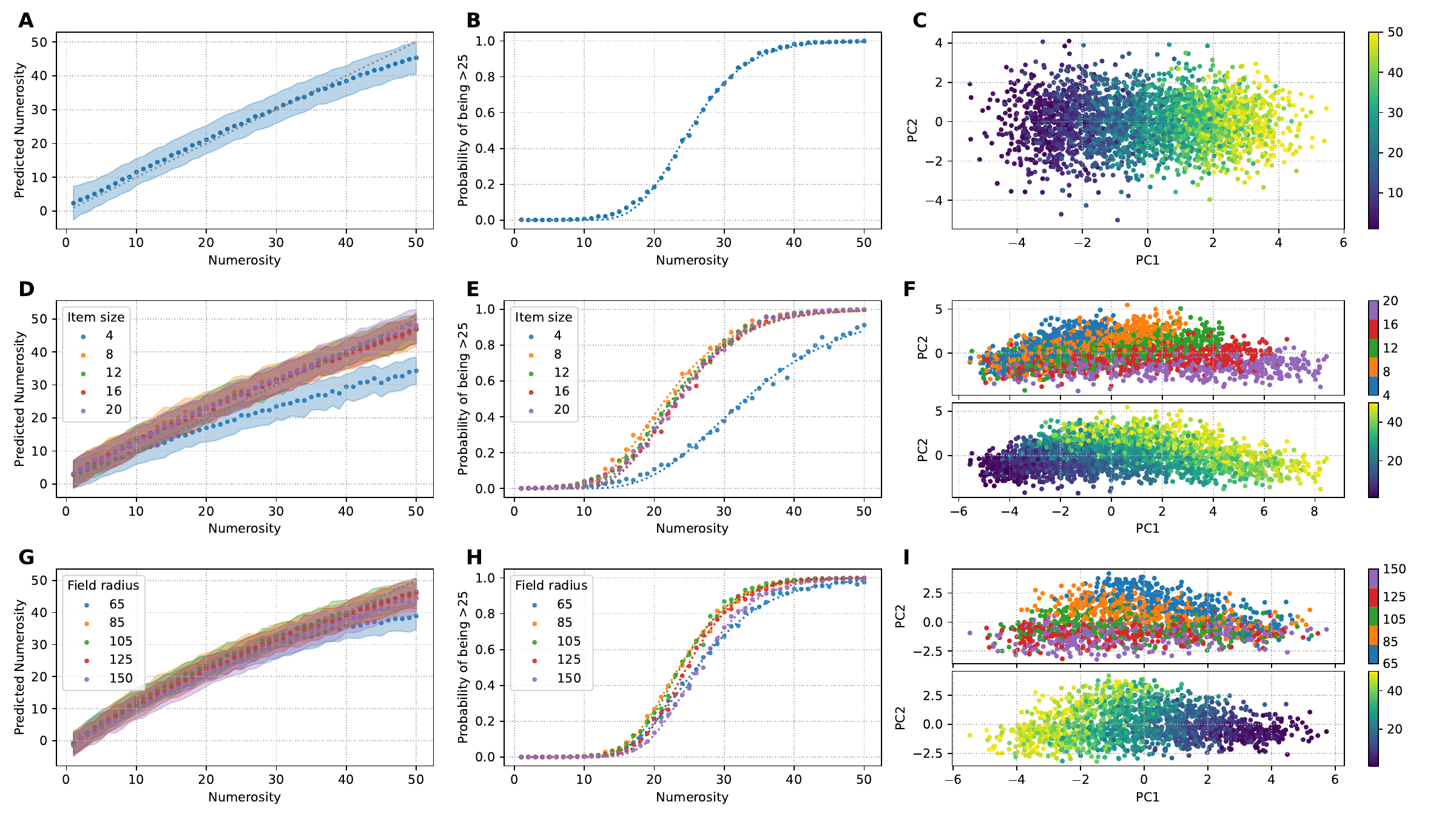}
    \caption{Analysis of the $\beta$-VAE-sup-50 model with retrained regressions. All panels are analogous to those described in Figure \ref{fig:bvae_retrained}. The generalization tests in panels D, E, G and H are conducted without re-training the model or the readouts. This allows comparison of the model's performance in numerosity estimation and discrimination tasks and the organization of its latent space under supervised training conditions and severe resource constraints.}
    \label{fig:bvae-sup-low_zeroshot}
\end{figure}

\clearpage

\subsection{Performance metrics}
\begin{table}[ht!]
\centering
\begin{tabular}{l|rrr|rrr|rrr|rrr}
\toprule
 & \multicolumn{3}{c|}{$\beta$-VAE-5000} 
 & \multicolumn{3}{c|}{$\beta$-VAE-50} 
 & \multicolumn{3}{c|}{$\beta$-VAE-sup-5000} 
 & \multicolumn{3}{c}{$\beta$-VAE-sup-50} \\
dataset & MAE & $w$ & $e^c$ & MAE & $w$ & $e^c$ & MAE & $w$ & $e^c$ & MAE & $w$ & $e^c$ \\
\midrule

test & 1.71 & 0.11 & 25.38 & 5.08 & 0.39 & 24.03 & 0.80 & 0.05 & 25.39 & 3.70 & 0.25 & 25.00 \\
&&&&&&&&&&&&\\
itemsize\_4 & 2.09 & 0.13 & 26.21 & 11.07 & 0.93 & 56.08 & 1.81 & 0.08 & 26.38 & 7.06 & 0.35 & 33.01 \\
itemsize\_8 & 1.76 & 0.12 & 25.07 & 6.07 & 0.66 & 23.96 & 1.77 & 0.07 & 23.95 & 4.24 & 0.35 & 21.66 \\
itemsize\_12 & 1.92 & 0.12 & 24.76 & 5.67 & 0.64 & 19.24 & 0.87 & 0.07 & 25.40 & 3.97 & 0.33 & 22.59 \\
itemsize\_16 & 2.23 & 0.14 & 25.48 & 5.99 & 0.63 & 17.45 & 1.11 & 0.06 & 26.29 & 3.74 & 0.32 & 23.37 \\
itemsize\_20 & 2.65 & 0.16 & 24.54 & 6.18 & 0.60 & 17.36 & 1.15 & 0.06 & 24.97 & 3.80 & 0.32 & 23.12 \\
&&&&&&&&&&&&\\
fieldradius\_65 & 1.78 & 0.11 & 25.87 & 4.54 & 0.34 & 25.05 & 1.59 & 0.08 & 26.05 & 4.41 & 0.30 & 26.03 \\
fieldradius\_85 & 2.22 & 0.10 & 25.22 & 4.08 & 0.29 & 22.88 & 1.33 & 0.08 & 25.19 & 3.75 & 0.26 & 23.54 \\
fieldradius\_105 & 2.15 & 0.10 & 25.14 & 3.83 & 0.28 & 23.43 & 1.22 & 0.08 & 25.38 & 3.54 & 0.24 & 23.74 \\
fieldradius\_125 & 2.08 & 0.11 & 25.28 & 3.67 & 0.25 & 25.00 & 1.28 & 0.08 & 25.47 & 3.38 & 0.23 & 24.64 \\
fieldradius\_150 & 2.30 & 0.11 & 25.50 & 4.18 & 0.27 & 27.12 & 1.39 & 0.09 & 25.28 & 3.61 & 0.23 & 26.26 \\

\bottomrule
\end{tabular}
\vspace{0.5cm}
\caption{Performance metrics obtained by $\beta$-VAE and $\beta$-VAE-sup models (at capacity 5000 and 50 nats) for numerosity estimation and discrimination tasks across different datasets (retrained readouts). MAE: Mean Absolute Error for numerosity estimation. $w$: Weber fraction for numerosity discrimination. $e^c$: internal criterion used for discrimination (see Section \ref{sec:met_behavioural_tests}). Results are shown after retraining the linear readouts on each dataset, with models' weights frozen. Each dataset corresponds either to the test distribution, to variations in item size, or to variations in field radius (see Section \ref{sec:met_robustness} for more details).}
\label{tab:metrics_retrained}
\end{table}

\clearpage
\subsection{Analysis of reconstructions of the $\beta$-VAE-5000 model}

Imposing capacity constraints on the $\beta$-VAE, while beneficial for studying rate-distortion trade-offs, introduces the risk that the model might prioritize optimizing the rate term of the loss function at the expense of the distortion term (i.e., reconstruction fidelity). To ensure that both terms are adequately optimized during training, we evaluated the reconstruction capability of the $\beta$-VAE-5000 model. Specifically, instead of directly analyzing the latent representations of the input images, we repeated the behavioral analyses (numerosity estimation and discrimination) on the \textit{reconstructions} of the input images. This allows us to verify that the model not only compresses the information effectively but also produces high-fidelity reconstructions.

\begin{figure}[ht!]
    \centering
    \includegraphics[width=1\linewidth]{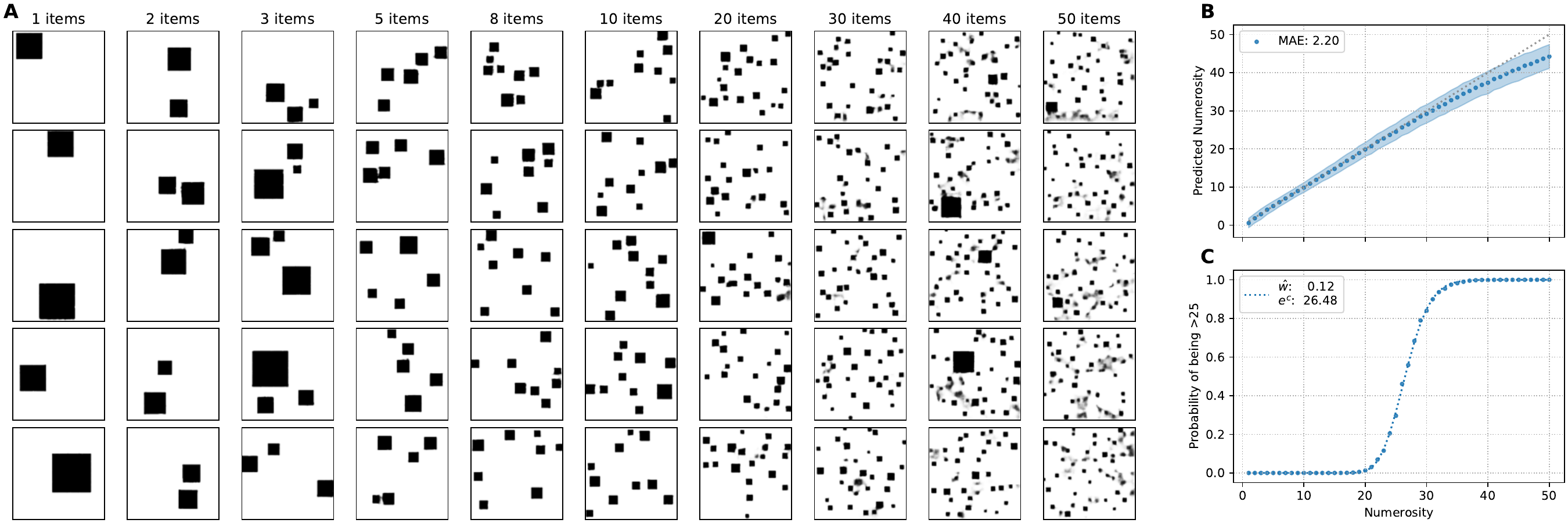}
    \caption{Analysis of reconstructions of the $\beta$-VAE-5000 model, with zero-shot evaluations. (A) Reconstruction samples (B) Zero-shot numerosity estimation test on latent representations of reconstructed samples (C) Zero-shot numerosity discrimination test on latent representations of reconstructed samples.} 
    \label{fig:decoding_unsup_highc_zeroshot}
\end{figure}

\clearpage

\subsection{Supplementary neural analyses}
To gain further insights into the neural coding strategies employed by the models, we performed various supplementary analyses on the latent representations. These analyses, including investigations of the relationship between neuron activity and numerosity, aim to characterize how numerosity information is encoded and organized within the models' latent space. The following figures present the results of these supplementary neural investigations.

\begin{figure}[ht!]
    \centering
    \includegraphics[width=\linewidth]{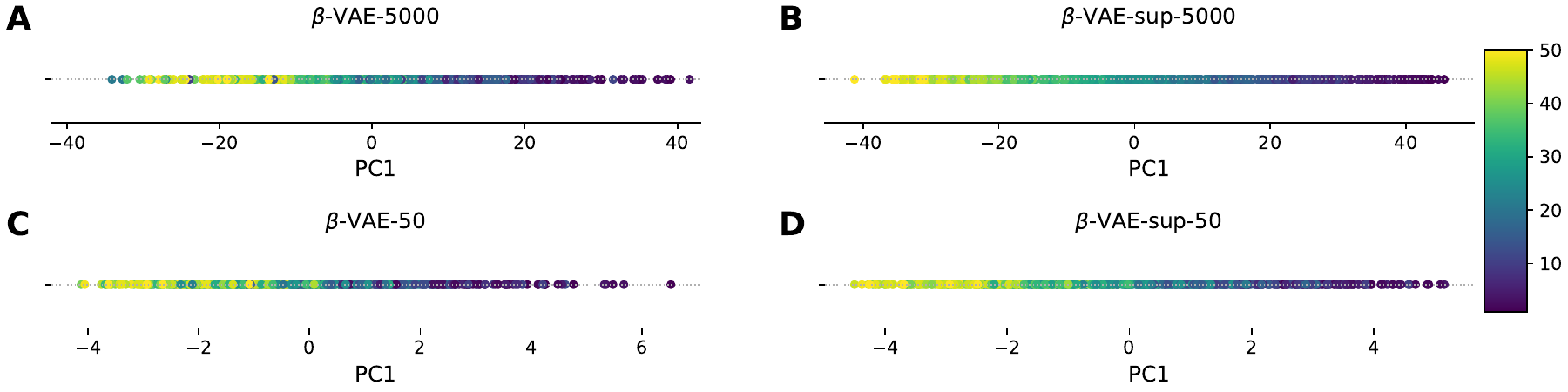}
    \caption{First principal component (PC1) of the latent space of each model. Data points are colored according to the numerosity of the input image. See the main text for further explanation.}
    \label{fig:na_numline}
\end{figure}

\begin{figure}[ht!]
    \centering
    \includegraphics[width=\linewidth]{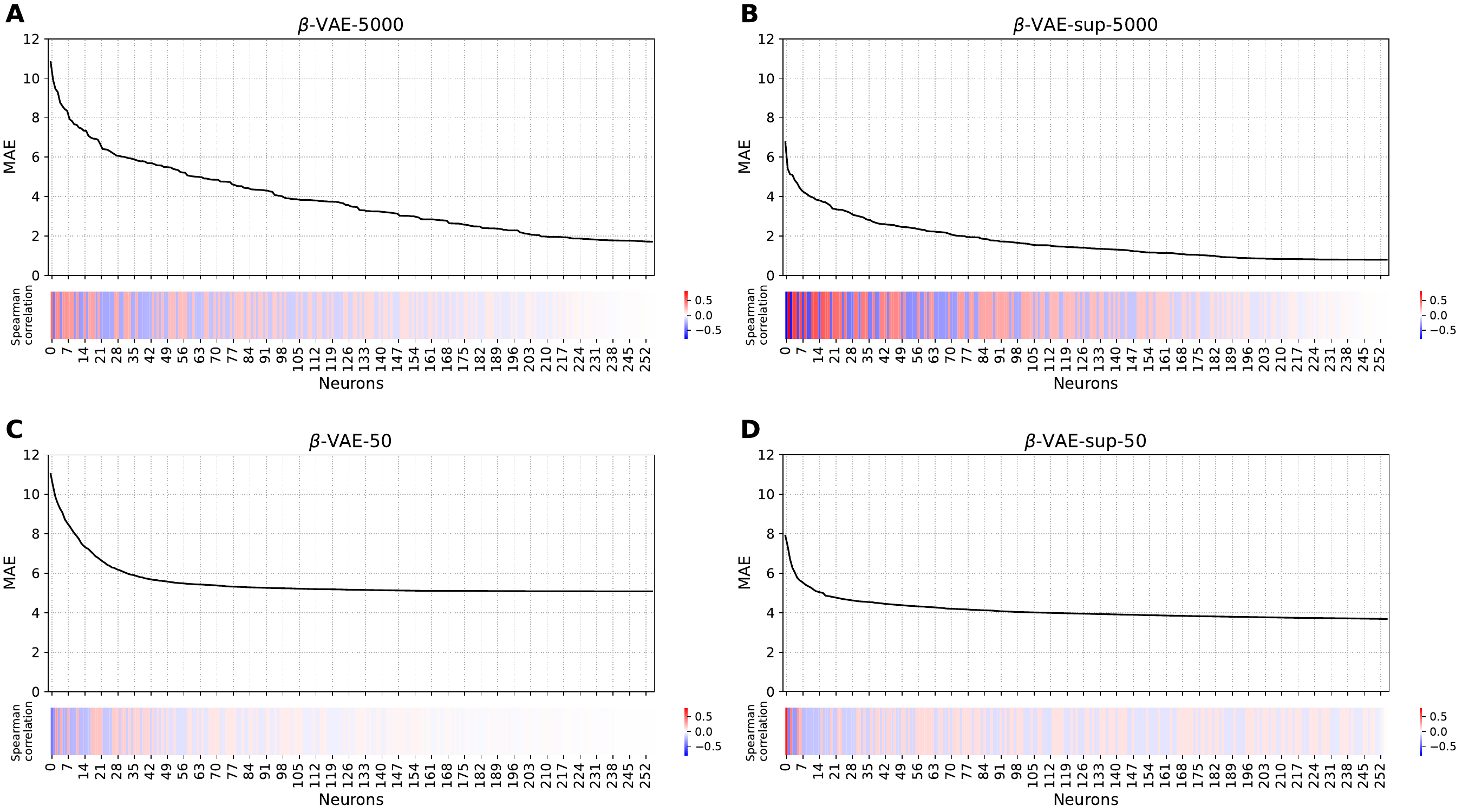}
    \caption{Redundant nature of the neural code. For each model, the plots show the MAE of 256 linear regressions trained to estimate numerosity from neural activity (top), using an increasing number of neurons, sorted by the correlation scores. The correlation score between neurons and numerosity is computed as spearman correlation (bottom). Note that having sorted the neurons by correlation scores implies that from panel to panel the order of the neurons may differ, despite having the same index. See the main text for further explanation.
 }
    \label{fig:na_corr_reg}
\end{figure}


\begin{figure}[ht!]
    \centering
    \includegraphics[width=\linewidth]{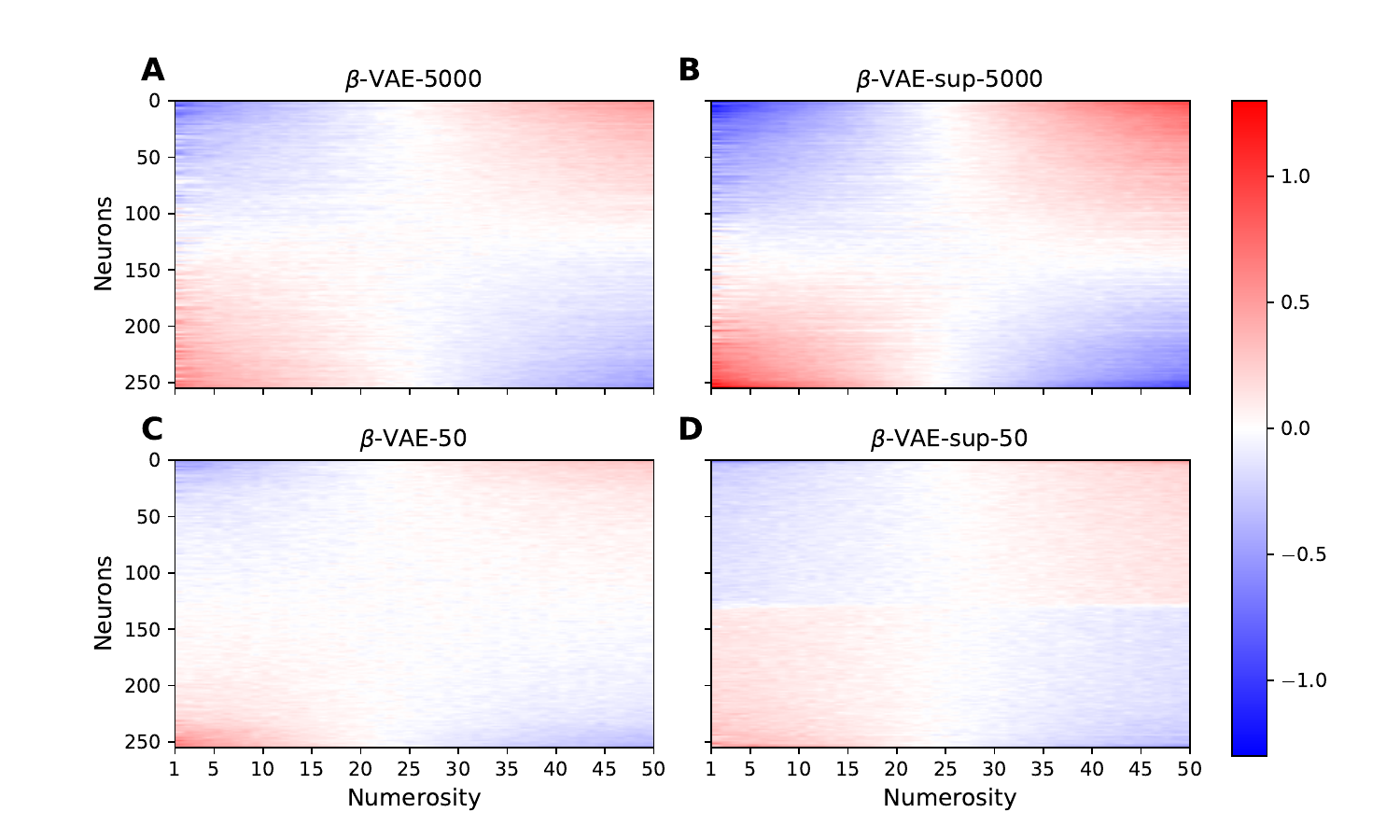}
    \caption{Heatmaps of average neuron activity as a function of numerosity. Each heatmap represents the average activity of each neuron in the latent space for images containing different numerosities. Activity levels have been standardized (z-scored). The color scale represents the standardized activity, with red indicating high positive activations and blue indicating negative activations. The majority of neurons span their full activity range to encode numerosities from 1 to 50, in keeping with a summation coding scheme. See the main text for further explanation.}
    \label{fig:na_heatmap}
\end{figure}


\clearpage

\subsection{Datasets used in generalization tests}
\begin{figure}[ht!]
    \centering
    \includegraphics[width=1\linewidth]{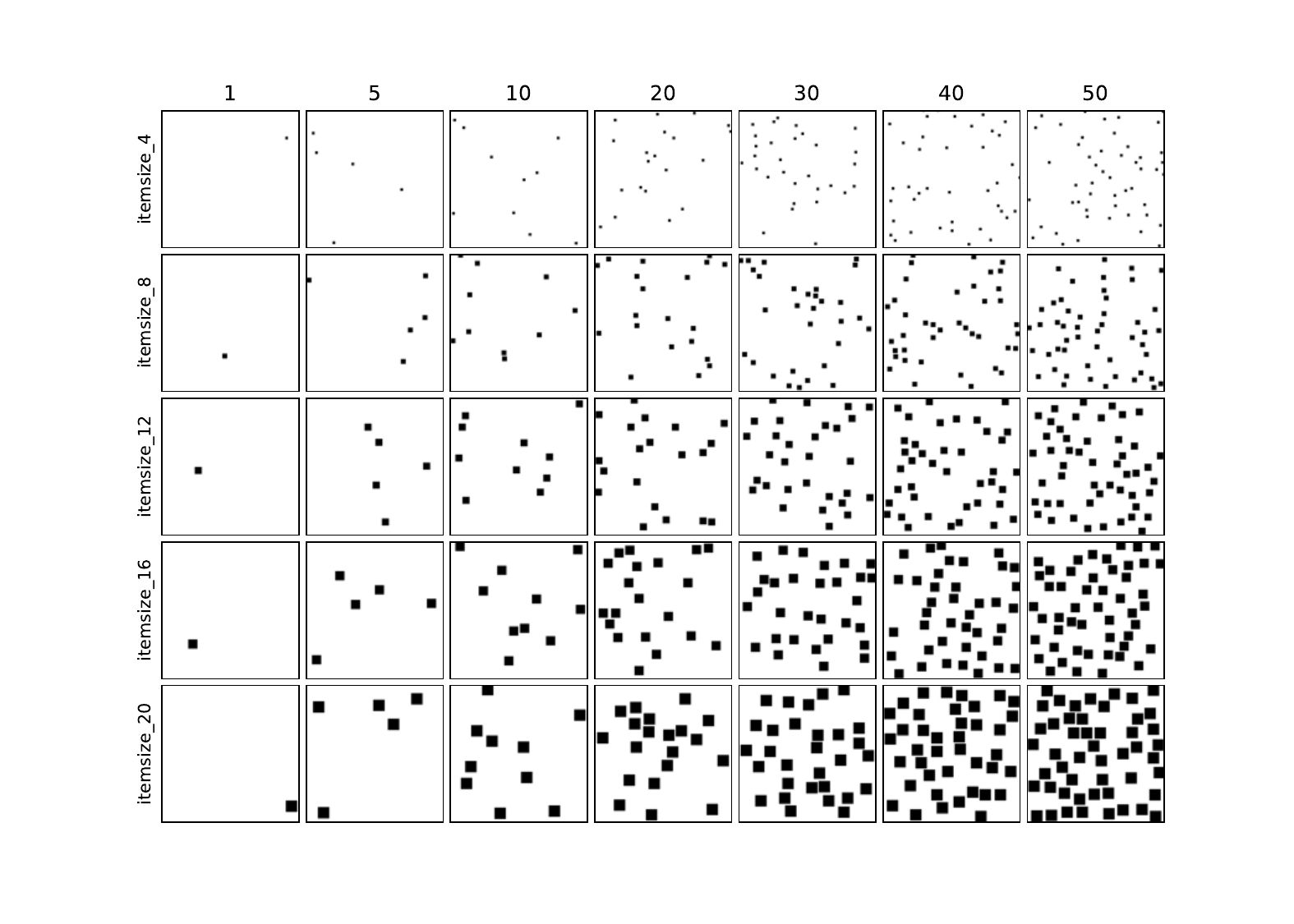}
    \caption{"Item size" dataset used for assessing generalization capabilities. Each subplot displays a representative sample with a specific combination of numerosity and item size. The size refers to the length measured in pixels of the edge of the squared items.}
    \label{fig:itemsize_dataset}
\end{figure}

\begin{figure}[ht!]
    \centering
    \includegraphics[width=1\linewidth]{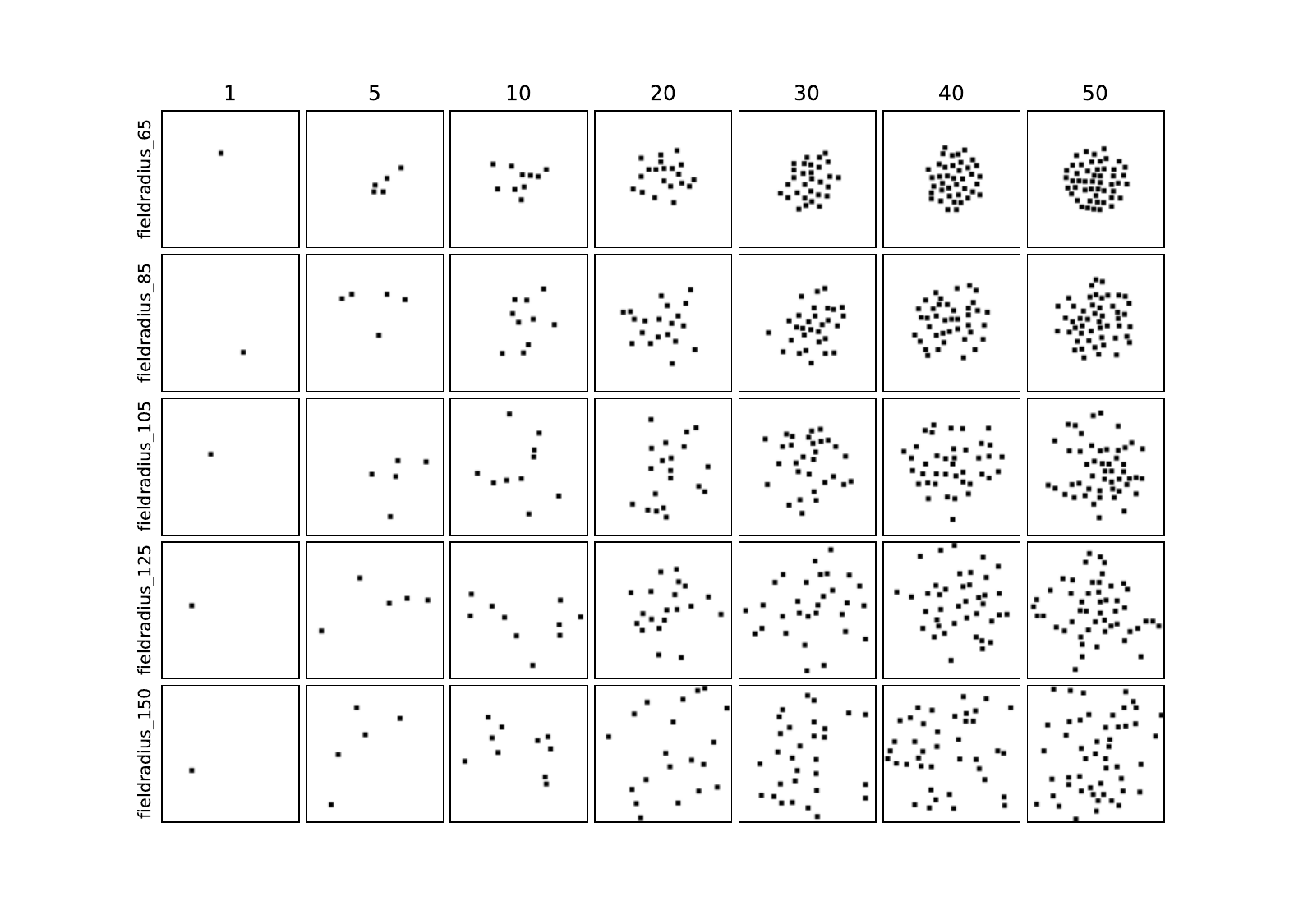}
    \caption{"Field radius" dataset used for assessing generalization capabilities. Each subplot displays a representative sample with a specific combination of numerosity and field radius. The field radius is measured in pixels from the center of the image and constrains the location of the items.}
    \label{fig:fieldradius_dataset}
\end{figure}

To evaluate the generalization capabilities of our models, we employed two novel datasets, "item size" and "field radius," which present conditions not encountered during training.  These datasets systematically vary the size and spatial distribution of items, allowing us to assess the robustness of the learned numerosity representations to changes in these visual attributes.  Figures \ref{fig:itemsize_dataset} and \ref{fig:fieldradius_dataset} illustrate representative samples from each dataset.

\clearpage
\subsection{Comparison to human data}
\begin{figure}[ht!]
    \centering
    \includegraphics[width=1\linewidth]{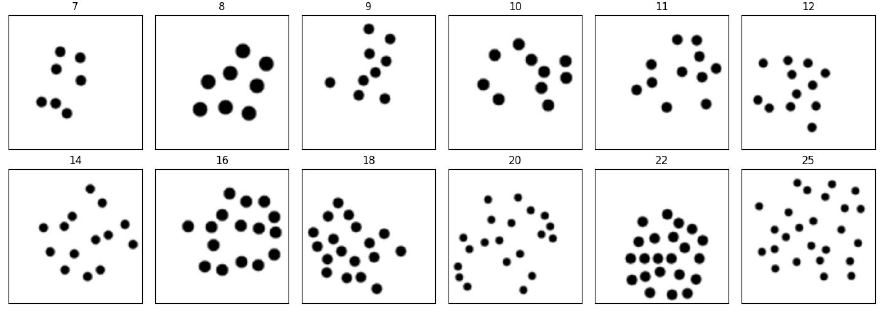}
    \caption{ Samples from "DeWind" dataset used to compare $\beta$-VAE models to human participants.}
    \label{fig:DeWind_dataset}
\end{figure}

To compare the performance of the artificial neural networks to that of human participants in numerosity discrimination tasks, Testolin et al. \cite{testolin2020visual} used the dataset reported in Figure \ref{fig:DeWind_dataset}, originally designed by DeWind et al.  \cite{DEWIND2015247}. The dataset was sampled from a multidimensional space defined by three orthogonal dimensions, representing the degrees of freedom used to generate all possible combinations of numerosity and non-numerical features in a visual display. Numerosity corresponds to the discrete number of dots in the image, Spacing jointly encodes for variations in field area and density of the dots, while Size jointly encodes for variations in dot surface area and total surface area.

In Figure \ref{fig:human_comp}D-F we compare our models to the human participants. From the figure we conclude that the $\beta$-VAE-5000 is slightly above the human performance. Here we repeat the same analyses for a model, the $\beta$-VAE-1000, that is more aligned to human data. 

\begin{figure}[ht!]
    \centering
    \includegraphics[width=1\linewidth]{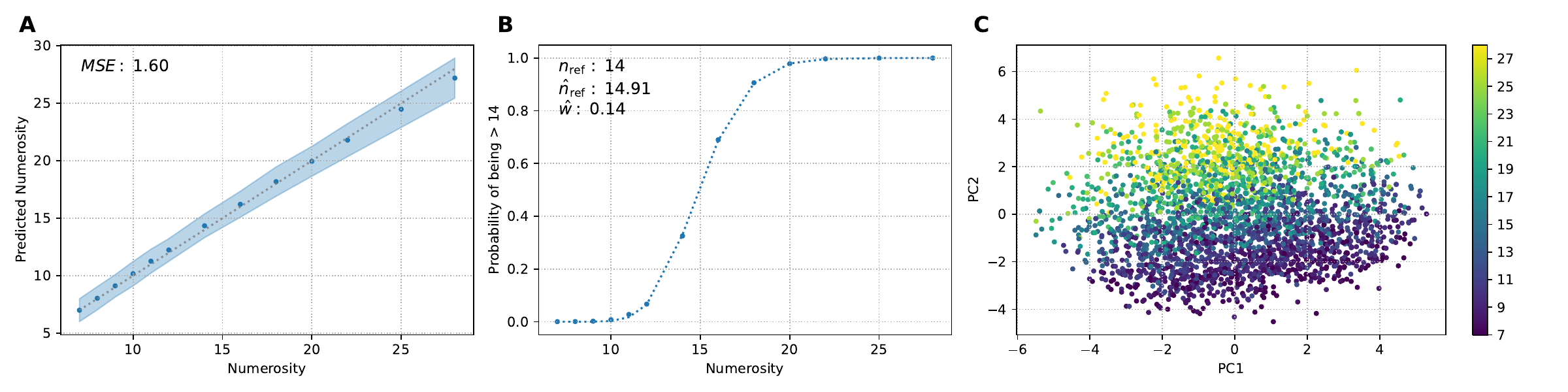}
    \caption{$\beta$-VAE-1000 model results on the DeWind dataset.}
    \label{fig:bvae-1000-dewind}
\end{figure}


\clearpage

\end{document}